\documentclass[manuscript,screen, nonacm]{acmart}

\usepackage{rotating}   
\usepackage{colortbl}   
\usepackage{xspace}
\newcommand{\zb}{\discretionary{}{}{}}   

\newcommand{\NPE}{\texttt{Null\zb Pointer\zb Exception}\xspace}
\newcommand{\ISE}{\texttt{Illegal\zb State\zb Exception}\xspace}
\newcommand{\IAE}{\texttt{Illegal\zb Argument\zb Exception}\xspace}
\newcommand{\CCE}{\texttt{Class\zb Cast\zb Exception}\xspace}
\newcommand{\AIOOBE}{\texttt{Array\zb Index\zb Out\zb Of\zb Bounds\zb Exception}\xspace}
\newcommand{\SIOOBE}{\texttt{String\zb Index\zb Out\zb Of\zb Bounds\zb Exception}\xspace}
\newcommand{\NFE}{\texttt{Number\zb Format\zb Exception}\xspace}

\AtBeginDocument{%
  }

\usepackage{booktabs}
\usepackage{multirow}
\usepackage{array}
\usepackage{graphicx}
\usepackage{xcolor}
\usepackage{amsmath}
\usepackage{listings}
\usepackage[htt]{hyphenat}
\usepackage{cleveref} 
\usepackage{placeins}
\usepackage{subcaption}
\usepackage[most]{tcolorbox}
\newtcolorbox{finding}[1]{colback=black!3,colframe=black!40,
  boxrule=0.5pt,left=6pt,right=6pt,top=4pt,bottom=4pt,
  title={\textbf{Finding #1}},fonttitle=\normalsize,coltitle=black,
  colbacktitle=black!3,titlerule=0pt}

\usepackage{float}
\usepackage{lstlinebgrd}
\definecolor{kw}{RGB}{0,0,180}
\definecolor{cmt}{RGB}{0,128,0}
\definecolor{str}{RGB}{163,21,21}
\definecolor{witnessbg}{RGB}{255,235,235}
\lstdefinestyle{java}{
  language=Java,
  basicstyle=\ttfamily\scriptsize,
  keywordstyle=\color{kw}\bfseries,
  commentstyle=\color{cmt}\itshape,
  stringstyle=\color{str},
  showstringspaces=false, columns=fullflexible, breaklines=true,
  frame=single, xleftmargin=1.4em, framexleftmargin=1.2em,
}

\newcommand{\runin}[1]{%
  \par\noindent\textit{#1}\nobreak\hspace{0.5em}%
}

\usepackage{needspace}
\usepackage{enumitem}

\setcopyright{none}
\renewcommand\footnotetextcopyrightpermission[1]{}

\newcommand{\boldrunin}[1]{%
  \par\noindent\textbf{#1}\nobreak\hspace{0.5em}%
}

\begin{document}
    
\title{RXSynth: A Seed-Grounded Corpus for Studying Java Runtime-Exception Bugs}

\author{Parnian Eslamloo}
\affiliation{%
  \institution{University of Saskatchewan}
  \city{Saskatoon}
  \country{Canada}}
\email{parnian.eslamloo@usask.ca}

\author{Banani Roy}
\affiliation{%
  \institution{University of Saskatchewan}
  \city{Saskatoon}
  \country{Canada}}
\email{banani.roy@usask.ca}

\renewcommand{\shortauthors}{Eslamloo et al.}

\begin{abstract}
Existing Java bug benchmarks provide too little method-level data for studying runtime-exception defect detection at scale. This also makes it difficult to determine how much evidence about a defect can be recovered from an isolated method body. Filtering Defects4J, Bugs.jar, Bears, and BugSwarm for in-scope unchecked runtime exceptions whose repair is contained within one method leaves only $170$ real bug/fix pairs across seven exception types.

We build RXSynth, a corpus of $4{,}275$ validated method-level buggy/fixed Java pairs generated from these real examples. We use seed-grounded few-shot prompting and screen candidate generators for alignment on structural metrics before production generation. Every released pair then passes a two-stage patch-anchored validator. The validator first localises the labelled risk and then checks whether the repair removes that same risk. A blinded human audit motivates the patch-anchored criterion, which we then evaluate through cross-judge comparison and execution. An independently generated JVM harness confirms patch-local validity for $84.2\%$ of a $500$-pair stratified sample. Each synthetic pair also records source-bug provenance for its generation exemplars, which enables provenance-disjoint evaluation. Fidelity analysis shows that RXSynth is structurally close to the fixed $157$-pair real reference set within the represented size range. In a fixed-generator \NPE ablation, the seed-grounded configuration, which combines real exemplars with size targets, achieves better structural alignment than unconditioned generation. However, larger exemplars are compressed, and incidental program context transfers less faithfully than the defect mechanism.

Using RXSynth, we find that no evaluated approach achieves strong and balanced isolated-method discrimination across the tested architectures, training sources, and inference settings. Supervised ROC-AUC remains about $0.55$--$0.64$, while the stronger zero-shot probes reach about $69\%$ accuracy but correctly classify only $41.2\%$ of repaired methods as non-defective. When the repair is visible, pair-discriminative signal transfers to unseen real defects. The fine-tuned encoders reach $70.4$--$74.8\%$ accuracy, while a parameter-free repair-growth rule, measured in tokens, reaches $84.7\%$ expected accuracy. On the $36$ pairs where the repair does not increase token count, the encoders fall to approximately chance. This is consistent with much of their aggregate performance being associated with repair-direction surface cues. Given both methods, GPT-5.5 and Gemini~3.1~Pro still exceed $94\%$ accuracy on the same residual subset, indicating that they can discriminate the pairs when the dominant repair-growth cue is absent.

These results position RXSynth as a resource for contrastive, patch-oriented, and context-aware research rather than an isolated-method detection leaderboard. More broadly, bug/fix evaluations should report directional surface baselines and provenance controls before high pairwise accuracy is interpreted as semantic defect understanding.
\end{abstract}



\maketitle
\section{Introduction}
\label{sec:intro}

Runtime exceptions are a persistent source of production failures in Java. Unlike checked exceptions, unchecked exceptions do not need to appear in method signatures and are not enforced by the compiler. \NPE has been reported as a leading cause of uncaught crashing failures in production Java~\cite{durieux2017dynamic}. Learning-based detection could help identify these defects before they cause failures. To study this possibility at method level, however, we need labelled examples containing both the defective method and the corresponding developer repair.

Existing Java bug benchmarks provide real defects and developer fixes, but they were not designed for controlled method-level studies. Defects4J~\cite{just2014defects4j}, Bugs.jar~\cite{saha2018bugs}, Bears~\cite{madeiral2019bears}, and BugSwarm~\cite{tomassi2019bugswarm} organise data around bugs or commits, whose patches may span several methods or files. Across these four benchmarks, we find $3{,}202$ bugs, of which $938$ fail with an in-scope unchecked runtime exception. Requiring the relevant repair to be contained within one method, and validating its connection to the labelled exception, leaves only $170$ buggy/fixed method pairs from $161$ source bugs across seven exception types. The pool is also imbalanced: \NPE accounts for $87$ pairs, while four exception types contribute fewer than $15$ pairs each.

This shortage makes the central question of our study difficult to answer: \emph{how much information about a runtime-exception defect can a model recover from an isolated method body, and what additional information becomes available when the developer's repair is visible?} Answering it requires enough data to compare models and training sources under controlled conditions, while retaining real defects for evaluation. Existing runtime-error resources do not directly provide this combination. RunBugRun~\cite{prenner2023runbugrun} offers large-scale buggy/fixed program pairs from programming-contest submissions, and ThrowBench~\cite{prenner2025throwbench} evaluates exception prediction on short executable programs. Both are useful, but neither provides a production-derived method-level corpus with traceable links to real defects. We therefore need additional examples without losing the connection to real bugs.

Synthetic generation can provide scale, but generating more code does not necessarily generate useful data. A method may be syntactically correct yet unlike production code in size, structure, or surrounding logic. Prior work shows that a mismatch between synthetic and target distributions can reduce downstream utility~\cite{alrashedy2024learning,ren2025few}. A generated pair may also contain the requested exception without reproducing the defect that its repair is supposed to address. We must therefore examine both \emph{fidelity} to real methods and \emph{validity} of the specific buggy/fixed relationship.

We address these requirements with \textbf{RXSynth}, a corpus of $4{,}275$ validated method-level buggy/fixed Java pairs. Its construction follows three principles: seed grounding, patch-anchored validation, and provenance tracking. First, we generate new pairs from real bug/fix exemplars rather than exception descriptions alone, and screen candidate generators for structural alignment before production generation. Second, we check whether each generated repair removes the labelled defect it is meant to fix. Third, we record which real source bugs supplied the generation exemplars. Because synthetic descendants of a real bug could otherwise be trained on while that bug is tested, these provenance records let us exclude them during evaluation.

Validation raises a question of its own: \emph{how can we tell whether a generated repair fixes the intended exception, rather than an unrelated risk elsewhere in the method?} We use a two-stage, patch-anchored validator. It first identifies the labelled runtime risk at the changed location, then checks whether the repair removes that same risk. The repaired method may still throw the labelled exception elsewhere. Divergent whole-method judgments in a blinded human audit motivate this criterion, which we further examine through cross-judge comparison and independent execution. On a stratified sample of $500$ released pairs, an independently generated JVM harness confirms patch-local validity for $84.2\%$. This is an execution confirmation rate, not a claim that every unconfirmed pair is invalid.

We also test whether the generated methods resemble their real exemplars. In a fixed-generator \NPE ablation, the seed-grounded configuration, which combines real exemplars with size targets, aligns more closely with the real seeds than unconditioned generation. The match is not complete: generated methods tend to compress exemplars beyond roughly $20$ statements and contain less incidental program context. RXSynth thus expands the available data while making its distributional limits explicit.

With this corpus, we examine two information settings. In the \emph{isolated-method} setting, a model receives one method body and predicts whether it is the defective or repaired member of a bug/fix pair. In the \emph{pair-discrimination} setting, it receives both versions and identifies the defective one. The second task makes the developer's change visible; it does not merely apply a different classifier to the same input.

Isolated-method discrimination remains weak under the evaluated conditions. Across provenance-controlled synthetic-to-real transfer and real-to-real training, supervised ROC-AUC is approximately $0.55$--$0.64$. Allowing provenance overlap raises Random Forest ROC-AUC to $0.83$. We treat this as a leakage diagnostic rather than evidence of transfer to unseen defects. Changing the inference regime does not resolve the difficulty: in zero-shot inference, GPT-5.5 and Gemini~3.1~Pro reach about $69\%$ accuracy but correctly identify only $41.2\%$ of repaired methods as non-defective, revealing a bias toward the defective class. The result therefore does not depend only on one model architecture or on synthetic training data. We interpret it as a limitation of the evaluated models and settings, not as proof that isolated methods contain no useful defect information.

In the repair-visible setting, synthetic-trained pair classifiers discriminate unseen real bug/fix pairs above chance: fine-tuned CodeBERT and CodeT5 reach $70.4\%$ and $74.8\%$ accuracy, respectively. But better pairwise accuracy does not necessarily mean better understanding of the defect. Repairs can introduce directional surface cues, such as added checks or statements, that distinguish the versions without identifying the underlying failure mechanism. Indeed, a parameter-free rule based only on repair growth, measured in tokens, reaches $84.7\%$ expected accuracy, exceeding both trained encoders. This raises a more specific question: \emph{are pair models recognising the defect, or are they exploiting properties of the repair?}

To test this, we examine the $36$ real pairs in which the repair does not increase token count. The usual pattern of positive repair growth is absent in this subset. CodeBERT and CodeT5 fall to approximately chance, with confidence intervals including $50\%$. In contrast, GPT-5.5 and Gemini~3.1~Pro remain above $94\%$ accuracy on the same pairs. The residual pairs are therefore still highly discriminable for these frontier models, although this does not by itself establish semantic reasoning. The encoder results are consistent with much of their aggregate pairwise performance being associated with repair-direction cues. \emph{High bug/fix accuracy should not automatically be interpreted as semantic defect understanding.}

Two implications follow. First, the weak isolated-method results motivate examining richer context (callers, callees, call sites, contracts, data-flow facts, and execution evidence) rather than only increasing model capacity while keeping the same input. Second, pairwise evaluations should include provenance controls, simple directional baselines, and residual analyses where dominant surface cues are absent. More broadly, synthetic scale, semantic validity, and distributional coverage are separate properties, so RXSynth extends scarce real data rather than replacing the diversity of production Java. It is intended mainly for contrastive, patch-oriented, and context-aware research, rather than as an isolated-method detection leaderboard.

\boldrunin{Research questions.}
RQ1 and RQ2 assess whether RXSynth provides a valid and sufficiently faithful basis for controlled study. RQ3 and RQ4 then examine what the two information settings reveal.
\begin{description}[leftmargin=0pt, labelindent=0pt]
  \item[RQ1 (Validity)] What fraction of generated pairs are valid under the patch-anchored criterion, and how reliable is that judgment? (Section~\ref{sec:validation})
  \item[RQ2 (Fidelity)] Does seed-grounded generation produce a corpus whose structural and lexical distributions match those of real runtime-exception bugs? (Section~\ref{sec:fidelity})
  \item[RQ3 (Isolated Method)] Can the defective member of a runtime-exception bug/fix pair be identified from its method body alone? (Section~\ref{sec:difficulty})
  \item[RQ4 (Pair Discrimination)] Can the defective member be identified when its repair is also visible, and what signal do models use to do so? (Section~\ref{sec:contrastive})
\end{description}

\boldrunin{Contributions.}
This work makes four contributions:
\begin{enumerate}[
    label=(\arabic*),
    wide=0pt,
    labelindent=0pt,
    itemsep=2pt
]
  \item \textbf{A production-grounded runtime-exception corpus.} We release $4{,}275$ validated synthetic method pairs across seven unchecked exception types, together with the $170$ real pairs, provenance records, excluded candidates, the zero-shot contrast corpus, and validation metadata.
  \item \textbf{A seed-grounded generation methodology.} We screen generators for structural alignment and generate from real bug/fix exemplars. A fixed-generator \NPE comparison evaluates the combined seed-grounded configuration, while fidelity analysis identifies its limits for larger and more context-rich methods.
  \item \textbf{A patch-anchored validation methodology.} We separate localisation of the labelled runtime risk from verification that the repair removes it, and evaluate this criterion through a blinded human audit, cross-judge comparison, and independent JVM execution.
  \item \textbf{A provenance-controlled study of two information settings.} Under the evaluated controls, no approach achieves strong and balanced isolated-method discrimination, while repair-visible discrimination transfers to unseen real defects. Simple repair-direction cues account for much of the trained pair-model performance, so pair accuracy requires repair-direction baselines and cue-controlled analysis before it can support claims about defect understanding.
\end{enumerate}
\section{Background and Related Work}
\label{sec:related}

This section positions RXSynth against prior work on Java runtime exceptions, real-world bug benchmarks, method-level prediction, learning-based defect detection, and synthetic bug generation. We focus on the aspects that motivate the corpus and the two information settings evaluated in Section~\ref{sec:usecase}.

\subsection{Java Runtime Exceptions and their Production Prevalence}
\label{sec:related:exceptions}

Java distinguishes checked exceptions, which are enforced through declaration or handling requirements, from unchecked exceptions that can arise without a compile-time warning. The seven types studied here are unchecked and need not appear in method signatures, making them a natural target for learning-based analysis. Their production relevance is well established. \texttt{NullPointerException} has been reported as a leading cause of uncaught crashing failures in production Java~\cite{durieux2017dynamic}. Empirical analyses of Java bugs also identify a small set of unchecked exception types as recurring failure modes~\cite{sobreira2018dissection,tomassi2019bugswarm,nguyen2019empirical}. We study \NPE, \IAE, \ISE, \CCE, \AIOOBE, \SIOOBE, and \NFE. Section~\ref{sec:runtime-def} defines the precise scope and exclusions used when extracting real examples.

\subsection{Java Bug Benchmarks}
\label{sec:related:benchmarks}
Our real examples come from Defects4J~\cite{just2014defects4j}, Bugs.jar~\cite{saha2018bugs}, Bears~\cite{madeiral2019bears}, and BugSwarm~\cite{tomassi2019bugswarm}. These resources provide real developer fixes together with failing tests, continuous-integration failures, or other failure metadata, making them suitable sources of production bug/fix evidence. Their native unit, however, is the bug or commit: a patch may span multiple methods or files, and the benchmarks are primarily structured for tasks such as fault localisation, failure reproduction, and automated program repair. They therefore do not directly provide the method-level buggy/fixed pairs required by our prediction tasks.

Other Java corpora illustrate complementary dataset designs. ManySStuBs4J~\cite{karampatsis2020often} provides large-scale single-statement bug data, while GitBug-Java~\cite{silva2024gitbug} collects more recent real bugs partly to reduce concerns about benchmark contamination in modern models. Neither is a seed source here. Our extraction instead starts from the four benchmarks above and requires the relevant repair to be contained within one method and causally connected to the observed runtime exception. Section~\ref{sec:gap} describes that reduction from benchmark bugs to the real seed pool used throughout the paper.

\subsection{Method-Level Binary Classification}
\label{sec:related:binary}

We distinguish two information settings built from the same buggy/fixed method pair. In the \emph{isolated-method setting}, a model receives one method body and predicts whether it is the defective or repaired member of its pair. In the \emph{pair-discrimination setting}, a model receives both versions in randomised order and identifies which member is defective. RQ3 and RQ4 study these settings separately because they answer different questions: RQ3 asks what can be predicted from the method body alone, whereas RQ4 asks what additional pair-discriminative information becomes available when the developer repair is also visible. Neither setting includes the surrounding class or execution context unless explicitly stated.

The isolated-method setting assigns $y=1$ to the defective method and $y=0$ to the repaired method. It differs from several adjacent tasks. Token-level localisation marks suspicious locations within code; program-level prediction operates on complete programs or execution outcomes; just-in-time defect prediction uses change and process information; and automated program repair produces a patch rather than a binary judgment. Our unit is the method because the real-seed extraction retains only repairs whose relevant change is contained within one method (Section~\ref{sec:gap}). The same underlying pair can then support either information setting without changing its defect label.

Buggy/fixed structure is common in software-engineering benchmarks, including Defects4J~\cite{just2014defects4j}, BugSwarm~\cite{tomassi2019bugswarm}, Bugs.jar~\cite{saha2018bugs}, and ManySStuBs4J~\cite{karampatsis2020often}. The important distinction is therefore not whether a repair exists in the dataset, but whether the prediction model is allowed to observe it. This distinction avoids treating the pair-discrimination setting of RQ4 as a leakage variant of RQ3: it is deliberately a different task with a different information set.

Method-level prediction also has a cautionary empirical history. Giger et al.~\cite{giger2012method} showed that change and source metrics can support bug prediction at method granularity, while Pascarella et al.~\cite{pascarella2020performance} later found that much of the apparent performance disappears under a more realistic time-aware evaluation. Their task concerns general method-level bug proneness rather than defective-versus-repaired runtime-exception pairs, but the result motivates our emphasis on evaluation design and on separating the information available to a model from the model architecture itself.

\subsection{Learning-Based Bug and Vulnerability Detection}
\label{sec:related:detection}
Our experiments span three model families: classical feature models, including Random Forests~\cite{breiman2001random} and XGBoost~\cite{chen2016xgboost}; sequence models trained from scratch, including TextCNN~\cite{kim2014textcnn} and BiLSTM~\cite{hochreiter1997lstm}; and pretrained code transformers, including CodeBERT~\cite{feng2020codebert}, CodeT5~\cite{wang2021codet5}, and CodeGPT~\cite{lu2021codexglue}, building on the transformer architecture~\cite{vaswani2017attention}. The more relevant prior result is not the distinction among these architectures but their sensitivity to evaluation conditions. Learned detectors that perform strongly on curated data can transfer poorly to more realistic target distributions~\cite{chakraborty2021deep}.

Risse and B\"ohme~\cite{risse2024uncovering} report detectors that can flag isolated functions as vulnerable yet have difficulty separating vulnerable functions from their fixes. Ding et al.~\cite{ding2024vulnerability} evaluate paired vulnerable and patched examples on which contemporary code models remain close to random, while prompted LLM studies likewise report weak function-level vulnerability discrimination~\cite{ullah2024llms,steenhoek2024comprehensive}. These results have been linked to spurious correlations, dataset bias, and overfitting: a model can learn features associated with a benchmark population rather than the property a repair changes.

That concern is particularly relevant to buggy/fixed pairs because developer repairs change both semantics and observable surface form. We therefore do not treat high pair accuracy as semantic evidence by default; Section~\ref{sec:contrastive} evaluates explicit surface baselines before interpreting the trained pair models. Context may matter as well. Bieber et al.~\cite{bieber2022static} show that whole-program behavioural prediction can benefit from descriptions of external resources, illustrating that source code alone does not necessarily contain all state relevant to runtime behaviour. Our experiments ask a narrower question: whether a related limitation appears for runtime exceptions at method granularity and whether changing the available information changes the result. Sections~\ref{sec:difficulty} and~\ref{sec:discussion} provide the empirical answer and its interpretation rather than assuming that explanation here.

\subsection{Synthetic Data Generation for Code Analysis}
\label{sec:related:synth}
Synthetic data is a common response to the scarcity of labelled bugs~\cite{allamanis2021selfsupervised,tian2024ssvd}. Rule-based mutation methods apply predefined or mined transformations to real code: DeepBugs~\cite{pradel2018deepbugs} injects identifier misuses, SemSeed~\cite{patra2021semseed} reuses edit scripts mined from real fixes, and BUGFARM~\cite{ibrahimzada2023bugfarm} injects bugs at locations chosen to challenge a detector. Learned approaches replace fixed mutation operators with neural generation; examples include CodeBERT-guided bug injection~\cite{richter2022learning} and BUGLAB's jointly learned generator and detector~\cite{allamanis2021selfsupervised}. LLM-based generation further allows complete buggy/fixed examples to be produced through prompting, but makes semantic validation and prompt sensitivity central concerns~\cite{tambon2025bugs,tony2024prompting,siddiq2024quality}.

For the present work, semantic validity alone is not sufficient. A generated method can contain a genuine defect while still being a poor training example if the surrounding synthetic distribution differs substantially from the real population. Alrashedy et al.~\cite{alrashedy2024learning} show that synthetic examples selected for proximity to real bugs can outperform a larger unfiltered synthetic corpus; in another transfer setting, using the full synthetic pretraining set can even reduce downstream performance. Ren et al.~\cite{ren2025few} similarly report stylistic and structural mismatch in zero-shot LLM generation and make distribution alignment an explicit objective in SynAlign. Together, these results motivate treating fidelity to the target distribution as a design concern rather than checking only whether generated code parses or contains the intended defect.

RXSynth follows that motivation in two stages. Candidate generators are screened for structural alignment to real seeds before the production campaign (Section~\ref{sec:generator-selection}), and Section~\ref{sec:fidelity} evaluates the resulting corpus against those seeds after generation. Unlike fixed mutation pipelines, production generation is conditioned on complete real buggy/fixed exemplars, while semantic validity is assessed separately by the pipeline of Section~\ref{sec:validation}. The resulting methodology therefore asks both whether each generated pair encodes the intended defect and whether the generated population resembles the production-derived examples it is intended to extend.

\subsection{Closest Existing Resources}
\label{sec:related:closest}
RunBugRun~\cite{prenner2023runbugrun} and ThrowBench~\cite{prenner2025throwbench} are the closest existing resources because both contain runtime-error examples with executable or paired evidence, but they operate at program rather than method granularity. Table~\ref{tab:closest-resources} summarises the distinction.

\begin{table}[tbp]
  \centering
  \caption{Closest runtime-error resources and their relation to RXSynth.}
  \label{tab:closest-resources}
  \small
  \begin{tabular}{@{}lp{0.13\linewidth}p{0.22\linewidth}p{0.24\linewidth}p{0.13\linewidth}@{}}
    \toprule
    Resource & Unit & Source & Primary role & Production Java \\
    \midrule
    RunBugRun
      & Program
      & Programming contests
      & Bug/fix training pairs
      & No \\
    ThrowBench
      & Program
      & Programming contests
      & Runtime-exception evaluation
      & No \\
    RXSynth
      & Method
      & Production-derived bug seeds
      & Bug/fix training and analysis
      & Seed-grounded \\
    \bottomrule
  \end{tabular}
\end{table}

RunBugRun is a multilingual collection of approximately $716{,}000$ buggy/fixed program pairs derived from Project CodeNet~\cite{puri2021codenet}, with roughly $13{,}000$ Java pairs. Its Java programs are short, self-contained contest solutions: the median is $18$ lines, fixes are restricted to at most six token changes, and \NPE accounts for only $2.6\%$ of its Java exceptions. It provides substantial training-scale paired data, but in a distribution different from production Java and at whole-program granularity.

ThrowBench contains $2{,}466$ short programs across four languages drawn from RunBugRun. It obtains runtime labels through execution and evaluates whether a model can predict which exception a program throws on a given input; it is an evaluation resource rather than a method-level training corpus. RXSynth targets the complementary setting: method-level buggy/fixed Java pairs grounded in examples extracted from production-code bug benchmarks, spanning seven unchecked exception types and designed to support both isolated-method and pair-discrimination experiments. The distinction is therefore one of granularity,
source population, and research task rather than corpus size alone.
\section{From Real Benchmarks to a Method-Level Seed Pool}
\label{sec:seeds}
\label{sec:gap}
RXSynth is grounded in real bug/fix examples, so we first construct a method-level seed pool. We extract this pool from four established Java bug benchmarks. This section defines the exception scope, describes the filters used to obtain method-level pairs, and reports the final seed distribution.

\subsection{Scope: Unchecked Runtime Exceptions}
\label{sec:runtime-def}

We define a runtime exception as an unchecked Java exception, that is, a subclass of \texttt{java.lang.RuntimeException} or \texttt{java.lang.Error}, thrown from production code during program execution. Unlike checked exceptions, unchecked exceptions need not appear in method signatures and are not required by the Java compiler to be caught or declared.

This definition excludes two categories of failures that may also surface at runtime. \emph{Checked exceptions}, such as \texttt{IOException} and \texttt{SQLException}, are excluded because Java requires potentially throwing code paths to declare or handle them. \emph{Test-framework artifacts}, such as \texttt{junit.framework.AssertionFailedError} and \texttt{org.junit.ComparisonFailure}, are also excluded because they represent test outcomes rather than production failures.

Two unchecked categories are within the definition but excluded from the target type set. \texttt{AssertionError} is omitted because most occurrences in the seed benchmarks arise from test-framework assertions; the remaining production cases are too few for per-class learning (see F1 in Section~\ref{sec:filter-rules}). Other \texttt{Error} subclasses, such as \texttt{OutOfMemoryError} and \texttt{StackOverflowError}, are excluded because they typically reflect resource or environmental conditions rather than method-local programmer errors and are also rare in the source benchmarks.

Within this scope, we study seven exception types: \texttt{NullPointerException} (NPE), \texttt{IllegalArgumentException} (IAE), \texttt{IllegalStateException} (ISE), \texttt{ClassCastException} (CCE), \texttt{ArrayIndexOutOfBoundsException} (AIOOBE), \texttt{StringIndexOutOfBoundsException} (SIOOBE), and \texttt{NumberFormatException} (NFE). Together they cover recurring unchecked failure modes in production Java~\cite{sobreira2018dissection,durieux2017dynamic}.

\subsection{Filtering Rules}
\label{sec:filter-rules}
We apply five filters uniformly across the four benchmarks when extracting seed pairs. F1 and F2 operate at the bug level; F3--F5 operate at the method-pair level.

\begin{description}[leftmargin=0pt, labelindent=0pt]
  \item[F1 (Test-framework exclusion).] Bugs whose only failure exception is \texttt{AssertionFailedError} or \texttt{ComparisonFailure} (or any subclass under the \texttt{junit.framework} or \texttt{org.junit} packages) are excluded. Bugs whose failure record contains both a JUnit artifact and a runtime exception are retained, treating the runtime exception as the root cause.

  \item[F2 (Checked-exception exclusion).] Bugs whose only failure is a checked exception are excluded, in accordance with the unchecked-only scope defined in Section~\ref{sec:runtime-def}. This includes the \texttt{IOException} family, \texttt{SQLException}, and project-specific checked exceptions such as \texttt{JsonMappingException} (which extends \texttt{IOException}).

  \item[F3 (Single-method containment).] Each retained pair must consist of a buggy and a fixed method whose change is fully contained within the body of one method or constructor declaration. Patches that span multiple methods, modify field declarations, or affect only documentation are dropped at extraction.

  \item[F4 (Intentionality filter for IAE and ISE).] For \texttt{IllegalArgumentException} and \texttt{IllegalStateException}, only pairs whose fix \emph{removes} or \emph{restructures} an unintended throw are retained. Pairs whose fix \emph{adds} a previously-missing throw, treating the absence of input or state validation as the bug, are excluded, as they would teach a classifier that defensive code is itself buggy.

  \item[F5 (Causal validation).] Each surviving pair is validated by a large language model (Claude Sonnet 4.5) prompted to confirm that (i)~the buggy method contains code that could directly cause the labelled exception at runtime, and (ii)~the fix specifically addresses that cause. Pairs are accepted at confidence $\geq 0.7$.
\end{description}

F5 applies only to the extraction and validation of the real seed pool. The separate semantic-validation procedure for generated pairs is described in Section~\ref{sec:validation}.

Applying the definition of Section~\ref{sec:runtime-def} with filters F1 and F2 to the four benchmarks yields the bug-level prevalence in Table~\ref{tab:bug-level-prevalence}: of 3{,}202 bugs pooled across the four benchmarks, 938 (29.3\%) fail with an in-scope runtime exception. Method-pair extraction (F3), intentionality filtering (F4), and causal validation (F5), together with per-pair quality validation, subsequently reduce these to the final merged corpus of 170 method pairs (Table~\ref{tab:per-source}).

\begin{table}[tbp] 
  \caption{Bug-level prevalence of in-scope runtime exceptions in the four seed benchmarks after filters F1 and F2 (Section~\ref{sec:filter-rules}).}
  \label{tab:bug-level-prevalence}
  \small
  \centering
  \begin{tabular}{lrrr}
    \toprule
    Benchmark & Total bugs & Runtime bugs & \% \\
    \midrule
    Defects4J & 854     & 142 & 16.6 \\
    Bears     & 251     & 60  & 23.9 \\
    BugSwarm  & 1{,}015 & 248 & 24.4 \\
    Bugs.jar  & 1{,}082 & 488 & 45.1 \\
    \midrule
    Pooled    & 3{,}202 & 938 & 29.3 \\
    \bottomrule
  \end{tabular}
\end{table}

\subsection{Distribution and Imbalance of the Final Seed Pool}
\label{sec:seeds:distribution}
After method-pair extraction (F3), intentionality filtering (F4), and causal validation (F5), the 938 in-scope runtime-exception bugs yield 170 validated method-level bug/fix pairs. Table~\ref{tab:per-source} reports their distribution across benchmarks and exception types. The pool is highly imbalanced: \NPE accounts for 87 of 170 pairs (51.2\%), while \NFE contributes 13, \CCE and \SIOOBE 9 each, and \ISE only 8. Four of the seven exception types therefore contribute fewer than 15 pairs, leaving too little data for controlled per-class training at the scale required by the models studied here. This scarcity and imbalance motivate the synthetic generation procedure of Section~\ref{sec:synth-generation}.

\begin{table}[tbp] 
  \centering
  \small
  \caption{Distribution of the $170$ real seed pairs by source benchmark and exception type. Abbreviations follow Section~\ref{sec:runtime-def}.}
  \label{tab:per-source}
  \begin{tabular}{lrrrrrrrr}
    \toprule
    Source     & NPE & IAE & AIOOBE & NFE & CCE & SIOOBE & ISE & Total \\
    \midrule
    Defects4J  &  20 &  13 &     13 &   6 &   4 &      8 &   2 &    66 \\
    Bugs.jar   &  32 &   7 &      3 &   0 &   2 &      0 &   5 &    49 \\
    Bears      &  14 &   3 &      2 &   7 &   2 &      1 &   1 &    30 \\
    BugSwarm   &  21 &   1 &      2 &   0 &   1 &      0 &   0 &    25 \\
    \midrule
    Total      &  87 &  24 &     20 &  13 &   9 &      9 &   8 &   170 \\
    \bottomrule
  \end{tabular}
\end{table}

\subsection{Fidelity Reference Set and Production Conditioning}
\label{sec:seeds:generation-pool}

The 170 pairs in Table~\ref{tab:per-source} constitute the complete real seed pool, but not all were used as few-shot exemplars. We did not impose a size cutoff a priori. During pilot generation, conditioning on unusually large methods revealed a systematic failure mode: the generated method often preserved the exception mechanism while compressing the surrounding program structure, dropping branches, loop nests, and auxiliary statements. Because the purpose of grounding is to transfer structural characteristics as well as the exception mechanism, seeds that the generator could not reproduce structurally were designated for exclusion from the fidelity reference set.

Method size differs substantially across exception types, so this exclusion was applied relative to each type rather than through a single global threshold. Thirteen upper-tail pairs were designated as reference exclusions: seven \NPE, three \AIOOBE, two \SIOOBE, and one \IAE. The remaining 157 pairs constitute the \emph{157-pair fidelity reference set}; all \NFE, \CCE, and \ISE seeds were retained. This fixed set is the reference population used in the RQ2 structural comparisons and represents the size range the fidelity analysis was designed to characterise. Section~\ref{sec:dist:structural} compares the released corpus with both this 157-pair reference and the complete 170-pair seed pool, while Section~\ref{sec:preferred-band} independently quantifies the size-compression effect.

The 13 reference exclusions remain part of the released real seed pool. An artifact audit of the production generation records shows that the fidelity reference set is not identical to the exemplars that actually conditioned generation. Two \AIOOBE pairs from \texttt{JacksonCore-19}, both among the 13 reference exclusions, were used as production exemplars and have 82 unique released descendants. Conversely, four \NPE source bugs inside the reference set did not condition any released pair. The generation records confirm their non-use, but the reason cannot be reconstructed because the historical per-type selection bins were not retained. At source-bug granularity, production conditioning covers 146 distinct source bugs. Exact method-pair attribution is ambiguous for \texttt{Lang-47}: its two real method pairs have the same recorded exemplar size (18 physical lines and 13 logical lines), and production provenance does not distinguish them. The 38 released pairs citing this source bug therefore cannot be assigned uniquely to one of its two methods. This ambiguity does not affect the source-bug unit used for leakage control in Section~\ref{sec:leave-seeds-out}. The released provenance records and audit tables document these differences.

\section{Corpus Construction}
\label{sec:synth-generation}
\label{sec:construction}
We build RXSynth in two generation campaigns. The first is unconditioned (zero-shot): the model receives no real bug/fix examples. The second uses seed-grounded prompting with the real pairs from Section~\ref{sec:seeds}, and only this second campaign forms the main corpus. We report the first campaign because its limitations motivate the grounded design and provide the comparison used in Section~\ref{sec:dist:confound}.

\subsection{Zero-Shot Generation: Initial Attempt and Observed Limitations}
\label{sec:gen:zeroshot}
Before adopting seed-grounded generation, we conducted a zero-shot campaign with GPT-4o. The model received no real bug/fix examples. Instead, each prompt specified the target exception type, a failure mechanism, a concrete trigger pattern, a realistic programming context, and constraints on the method-level output format. The mechanism--trigger--context scaffolding was intended to promote semantic variety without relying on real exemplars.

Both generation campaigns applied syntactic and deduplication checks, although their archived configurations were not fully identical (Section~\ref{sec:gen:qc}). To compare their semantic validity under the criterion used in this paper, we evaluated a stratified sample of $150$ zero-shot pairs per exception type ($1{,}050$ pairs) with the two-stage patch-anchored validator of Section~\ref{sec:validation}. Aggregate zero-shot validity is $85.6\%$, comparable to the $83.9\%$ first-pass acceptance rate across the $4{,}900$ seed-grounded candidates. The main problem with zero-shot generation is therefore not a uniform loss of semantic validity.

Instead, two limitations motivated the grounded redesign. First, the generated methods are systematically shorter and structurally simpler than the real production bugs, clustering in a narrow complexity range despite variation in the prompts. Section~\ref{sec:distribution} quantifies this mismatch, and Section~\ref{sec:dist:confound} later isolates the effect of grounding under a fixed generator.

Second, zero-shot generation fails strongly for \texttt{IllegalArgumentException}. Only $36.0\%$ of the audited IAE pairs satisfy the patch-anchored criterion, compared with substantially higher rates for the other types. We ran a follow-up probe with a revised template that explicitly required the buggy method to throw and prohibited ``fixes'' that merely introduce a new throw. The follow-up implementation also disabled a buggy-already-validated gate, which rejected candidates whose buggy method contained two or more explicit argument-validation constructs. The follow-up configuration yields $58.7\%$ validity ($n=150$), compared with $36.0\%$ in the baseline audit, while seed-grounded generation reaches $81.1\%$.

The retained generation code identifies the baseline IAE validation-density gate and shows that the follow-up probe disabled it. Because the prompt template and the gate setting changed together, the $22.7$-percentage-point difference is a configuration-level comparison, not an isolated estimate of the prompt revision's effect. More importantly, the semantic failure mode remains. Of the $62$ invalid pairs produced under the follow-up configuration, $59$ still show the pattern explicitly prohibited by the revised prompt: the buggy method does not throw, while the fixed method introduces the exception. Thus, even under the revised instructions, the IAE semantic inversion remains and continues to motivate the grounded redesign.

These results motivated conditioning generation directly on real buggy/fixed exemplars. We retain the $7{,}000$ in-scope zero-shot pairs as a separate contrast corpus. They are not part of the main RXSynth corpus and are not used in the training or difficulty experiments.

\subsection{Generator Selection by Distributional Alignment}
\label{sec:generator-selection}

The seed-grounded protocol of Section~\ref{sec:gen:fewshot} fixes the prompt template and exemplars, leaving the generator as the main design choice. Before production, we screened six candidate generators under identical prompts, exemplars, and per-quartile budgets. Screening used only \NPE seeds: the $76$ \NPE seeds that parsed successfully and fell inside the quartile range used for screening, excluding the upper-tail seeds. We divided $76$ seeds spanning the generation size range into four equal complexity quartiles ($n=19$ each), and each candidate generated $19$ pairs per quartile. Where the serving API exposed sampling controls, we used temperature $0.9$ and top-$p$ $0.9$; the two Claude runs used provider defaults. Output was capped at $4{,}096$ tokens for all models.

Within each quartile, we compared the generated pairs with the corresponding real seeds on six structural metrics. For each metric, we computed the absolute Cliff's delta $|\delta|$ and averaged the six values to obtain a quartile score. Table~\ref{tab:generator-distribution-alignment} reports the four quartile scores and their mean. Claude Opus~4.7 has the lowest mean $|\delta|$ ($0.134$) and the best alignment in three of the four quartiles. GPT-5 performs best in the smallest-method quartile but diverges increasingly as method size grows. We therefore selected Claude Opus~4.7 for production generation. Per-metric results are provided in the replication package.

Selecting the generator by structural alignment creates a potential circularity because Section~\ref{sec:fidelity} later evaluates structural fidelity. The screening was limited to \NPE and a small candidate-generation sample, so six exception types contributed nothing to model selection. Moreover, Section~\ref{sec:dist:confound} holds the generator fixed and shows that the seed-grounded configuration improves alignment for \NPE. The full fidelity analysis therefore tests substantially more data than were used during generator selection.

\begin{table}[tbp]
  \centering
  \caption{Structural alignment of six candidate generators with real \NPE seeds: mean $|\delta|$ per complexity quartile (lower is better).}
  \label{tab:generator-distribution-alignment}
  \small
  \begin{tabular}{lrrrrrr}
    \toprule
    Generator & Q1 & Q2 & Q3 & Q4 & Mean & Rank \\
    \midrule
    Claude Opus   & 0.208 & \textbf{0.093} & \textbf{0.124} & \textbf{0.112} & \textbf{0.134} & 1 \\
    DeepSeek      & 0.146 & 0.210 & 0.300 & 0.167 & 0.206 & 2 \\
    GPT-4o        & 0.179 & 0.158 & 0.227 & 0.260 & 0.206 & 3 \\
    Claude Sonnet & 0.231 & 0.197 & 0.322 & 0.161 & 0.228 & 4 \\
    Qwen          & 0.201 & 0.473 & 0.359 & 0.224 & 0.314 & 5 \\
    GPT-5         & \textbf{0.110} & 0.316 & 0.470 & 0.384 & 0.320 & 6 \\
    \bottomrule
    \multicolumn{7}{l}{\footnotesize Mean over six structural metrics;
      per-metric detail in the replication package.} \\
  \end{tabular}
\end{table}

\subsection{Seed-Grounded Few-Shot Generation}
\label{sec:gen:fewshot}
Generation is conditioned on real exemplars rather than on a written description of each exception. Each prompt presents one or two real buggy/fixed pairs of the target type and asks the model to produce a new pair that follows the pattern in those exemplars. The prompt also includes a size target derived from the examples. We do not describe the intended bug pattern in prose; the exemplar pairs and their diffs provide that signal. This design reduces the role of hand-written semantic hints in shaping the generated bug pattern.

\runin{Prompt construction.}
Each exemplar contains three parts: the buggy method as it appeared in production, the fixed method, and a unified diff isolating the lines the developer changed. We compute the diff automatically with two lines of surrounding context. The prompt treats this change as the main pattern signal and instructs the model to follow the diff rather than the surrounding method body. This focuses generation on the defect itself, which may occupy only a few lines of a much larger method.

The prompt then imposes three requirements. The new pair must reproduce the same \emph{kind} of mistake visible in the exemplar diffs (for example, a flipped predicate, off-by-one error, missing guard, or wrong type); the fix must have the same \emph{shape} as the exemplar diffs; and the generated code must be genuinely new, with different identifiers, domain, and logic. Additional hard constraints require identical signatures for the buggy and fixed methods, forbid imports and helper declarations, and forbid comments that refer to the bug or fix.

Figure~\ref{fig:example-pair} shows a representative released pair and the local repair pattern captured by the corpus.

\begin{figure*}[t]
\centering
\definecolor{fixbg}{RGB}{225,245,225}
\begin{minipage}[t]{0.49\textwidth}
{\footnotesize\textbf{(a) Buggy}\hfill\emph{throws \NPE}}
\begin{lstlisting}[style=java, basicstyle=\ttfamily\fontsize{6.5}{7.5}\selectfont,
                   breaklines=true, escapeinside={(*@}{@*)}]
public String formatRecipientHeader(Recipient recipient, String defaultDomain, int maxLength) {
    StringBuilder header = new StringBuilder();
    String displayName = recipient.getDisplayName();
    String address = recipient.getAddress();
    if (address == null) {
        address = recipient.getUsername() + "@" + defaultDomain;
    }
    if (displayName != null && displayName.length() > 0) {
        header.append("\"").append(displayName).append("\" ");
    }
    header.append("<").append(address).append(">");
    String result = header.toString();
    if (result.length() > maxLength) {
        result = result.substring(0, maxLength - 3) + "...";
    }
    String priority = recipient.getPriority();
(*@\colorbox{witnessbg}{\parbox{\linewidth}{\ttfamily\ \ \ \ if (priority.equalsIgnoreCase("HIGH")) \{}}@*)
        result = "!! " + result;
    }
    return result;
}
\end{lstlisting}
\end{minipage}\hfill
\begin{minipage}[t]{0.49\textwidth}
{\footnotesize\textbf{(b) Fixed}\hfill\emph{literal is the receiver}}
\begin{lstlisting}[style=java, basicstyle=\ttfamily\fontsize{6.5}{7.5}\selectfont,
                   breaklines=true, escapeinside={(*@}{@*)}]
public String formatRecipientHeader(Recipient recipient, String defaultDomain, int maxLength) {
    StringBuilder header = new StringBuilder();
    String displayName = recipient.getDisplayName();
    String address = recipient.getAddress();
    if (address == null) {
        address = recipient.getUsername() + "@" + defaultDomain;
    }
    if (displayName != null && displayName.length() > 0) {
        header.append("\"").append(displayName).append("\" ");
    }
    header.append("<").append(address).append(">");
    String result = header.toString();
    if (result.length() > maxLength) {
        result = result.substring(0, maxLength - 3) + "...";
    }
    String priority = recipient.getPriority();
(*@\colorbox{fixbg}{\parbox{\linewidth}{\ttfamily\ \ \ \ if ("HIGH".equalsIgnoreCase(priority)) \{}}@*)
        result = "!! " + result;
    }
    return result;
}
\end{lstlisting}
\end{minipage}
\caption{A representative released \NPE pair (\texttt{fewshot\_NullPointerException\_fe543ce4}). The buggy method dereferences a possibly null \texttt{priority} (red); the fix makes the string literal the receiver (green).}
\label{fig:example-pair}
\end{figure*}

\runin{Exemplar selection.}
Each prompt uses one or two exemplars. For $k=2$, we enumerate all $\binom{N}{2}$ distinct seed pairs of the target type, shuffle them, and exhaust the queue before cycling, so every seed pair is used before any pair is reused. For $k=1$, and for types too small to support pair enumeration, we use LOC-stratified rotation across size bands while excluding recently used seeds to prevent short-term repetition. Of the released pairs, $3{,}157$ ($k=2$) were generated under pair-queue selection and $1{,}118$ ($k=1$) under stratified rotation.

\runin{Size target.}
Because ungrounded generation collapses toward short methods (Section~\ref{sec:construction}), most prompts carry an explicit size target derived from their exemplars ($3{,}816$ of the $4{,}275$ released pairs). The target is an approximate logical-line count with a tolerance band of $\pm40\%$. When two exemplars differ in size, the prompt instructs the model to aim between them and not to default to a short method. The size target is stated to be as important as the bug pattern. Section~\ref{sec:preferred-band} shows that the generator nonetheless reproduces exemplar size faithfully only below roughly $20$ statements and compresses larger exemplars. This undershoot therefore occurs despite an explicit size target.

\runin{Generation settings.}
All released pairs were generated with Claude Opus~4.7, selected by the distributional screening of \Cref{sec:generator-selection}. The model exposes no temperature, top-$p$, or top-$k$ controls, so generation used its default sampling with an output limit of $4{,}096$ tokens. Every accepted method is canonicalised by stripping comments before it enters the corpus. This is applied uniformly, regardless of comment content, and removes any possibility of label leakage through a comment such as \texttt{// added null check} without
relying on hand-written rules to decide what counts as a leak.

\subsection{Generation-Time Quality Control and Deduplication}
\label{sec:gen:qc}
Both generation campaigns applied syntax, novelty, and deduplication checks. However, their archived configurations were not identical, so the differences between these campaigns cannot be attributed solely to grounding or generator choice. The following checks describe the shared quality-control objectives, while Section~\ref{sec:gen:zeroshot} identifies the IAE gate difference.

Every generated pair first passes a syntactic gate. The output must contain exactly two fenced Java blocks: one buggy method and one fixed method with matching signatures. We remove import and class-wrapper boilerplate, and neither block may contain a class, interface, or enum declaration. Both methods are then wrapped in a dummy class and parsed with \texttt{javalang}. Malformed outputs and outputs containing additional methods are rejected. We later use the same \texttt{javalang} statement-node definition to measure logical lines in Section~\ref{sec:fidelity}.

Pairs that pass the syntactic gate are then checked for novelty and non-triviality. Each generated method must remain below the token-$n$-gram similarity threshold against every seed of the target exception type. The buggy and fixed versions must also remain below the within-pair similarity threshold, which rejects trivial or effectively empty patches. Generated methods must additionally satisfy the size constraints of their generation configuration. We remove near-duplicate generated pairs using normalised token 5-gram Jaccard similarity and then apply NiCad~\cite{cordy2011nicad} clone detection with rename-blind matching.

Most rejections come from the similarity checks, not from malformed Java. The two most common causes are insufficient distance from a seed and near-identity between the buggy and fixed methods. Parse errors, stray type declarations, multi-method outputs, and other syntactic failures account for a smaller share. Most filtering therefore removes examples that are too similar to the seeds or contain trivial patches. Per-attempt acceptance and rejection logs, including per-cause counts, are included in the reproduction package.

The two generation arms share broad syntax, structural-quality, and deduplication objectives, but their historical gate configurations were not fully identical. They differ in their grounding and generating model, a confound examined directly in Section~\ref{sec:dist:confound}. For the seed-grounded arm, per-type generation targets follow seed-pool prevalence, with a floor of $500$ and a cap of $1{,}000$ candidates per exception type. After these generation-time checks, $4{,}900$ candidate pairs proceeded to semantic validation in Section~\ref{sec:validation}.

\section{Validation (RQ1)}
\label{sec:validation}
Generation-time checks establish syntax, novelty, and non-triviality, but they do not establish semantic validity. A generated pair is useful only if it genuinely represents the exception it is labelled with. We therefore use a patch-anchored criterion motivated by a blinded human audit and evaluate it with three LLM judges from two model families. We then apply the criterion to all $4{,}900$ candidates with a predefined repair-and-readmit procedure and corroborate it through JVM execution. Because LLM-as-a-judge methods can be biased and task dependent~\cite{zheng2023judging,wang2024large,huang2025empirical,tong2024codejudge}, no single model verdict is treated as ground truth.

\subsection{Patch-Anchored Validity Criterion}
\label{sec:val:semantic}

A generated pair is valid only if it satisfies a patch-local criterion: \textbf{the buggy method throws the labelled exception at the site addressed by the patch, and the fixed method removes that specific risk at that site.} An unrelated occurrence of the same exception elsewhere in the method is not enough. Likewise, the repaired method does not need to be globally incapable of throwing that exception. The criterion asks whether the pair captures the particular defect addressed by the patch, not whether the whole method is universally safe.

A single yes/no judgment is not enough to establish this. The validator must first identify the relevant throw site and trigger, and then determine whether the repair removes that same risk. We therefore use two stages.

\runin{Stage 1: localise the risk and trigger.}
The judge receives the buggy method, the fixed method, and the target exception type $T$. It first summarises the patch and then checks whether the buggy version can throw $T$ at the changed location. If so, it identifies the exact throwing expression and a concrete \emph{trigger}: an argument value or object state that makes that expression throw. The prompt excludes unrelated $T$-risks elsewhere in the method. If the patch does not address $T$, the judge must return \texttt{can\_throw\_T = false} rather than choose a different risk. We assume single-threaded execution without reflection or caller-side validation, so the trigger must describe a concrete state that an ordinary caller could supply.

\runin{Stage 2: verify the repair on the same trigger.}
Stage~2 receives the site and trigger identified in Stage~1 together with both methods. It asks whether the same trigger still causes $T$ at that site in the fixed version. The scope remains patch-local. An explicit throw or another path to $T$ elsewhere in the repaired method does not invalidate the pair. The repair is valid if it guards, narrows, or otherwise removes the original triggering condition so that the Stage~1 trigger no longer reaches $T$ at the patched site.

A pair is accepted only when Stage~1 confirms the labelled exception at the patched site and Stage~2 confirms that the repair removes that risk. If Stage~1 finds no patch-local $T$-risk, the pair is rejected without running Stage~2. This differs from whole-method safety, which asks whether the repaired method can throw $T$ anywhere. Section~\ref{sec:val:robustness} shows that this distinction materially changes the resulting judgments.

\runin{Judge and record.}
Bulk validation uses Gemini~2.5~Flash for throughput; Section~\ref{sec:val:robustness} compares its decisions with two stronger judges. Each stage returns a structured record containing the patch summary, throwing site, trigger, verdict, and confidence. JSON output is requested, with a plain-text parser as fallback. The append-only pipeline is keyed by sample identifier, making validation resumable and preventing duplicate judgments within a validation pass.

\subsection{Why Patch Anchoring Is Necessary}
\label{sec:val:robustness}
The patch-anchored criterion was not chosen in advance. It emerged from a blinded audit of $170$ generated pairs spanning all seven exception types and the size-conditioned generation runs. Two Java-experienced annotators scored each pair independently, and the same sample was also evaluated by whole-method automated judges. The audit revealed a clear asymmetry. The annotators agreed on whether the buggy method could throw the labelled exception in $91\%$ of pairs, but agreed on whether the repair resolved it in only $63\%$ ($\kappa=0.30$). Their pair-level acceptance rates also differed substantially, at $57.1\%$ and $28.2\%$. Whole-method automated judging showed a similar spread (Table~\ref{tab:rubric-rates}).

Post-audit analysis showed where the disagreement came from. The annotators had interpreted the task as asking whether the repaired method could throw the labelled exception \emph{anywhere}. The whole-method automated rubric asked the same question. This is an existential safety question, not the label semantics intended for RXSynth. A repair can remove the defect addressed by the patch while leaving an intentional throw or another path to the same exception on a different input. Rejecting such a pair would confuse patch-local correctness with whole-method exception freedom. The patch-anchored criterion therefore asks whether the labelled exception occurs at the site addressed by the patch and whether the repair removes that specific risk.

Re-evaluating the audit sample under the patch-anchored criterion produces much closer decisions across the three judges. Acceptance ranges from $82.9\%$ to $86.5\%$ across Gemini~2.5~Flash, Gemini~2.5~Pro, and GPT-5.5 (Table~\ref{tab:rubric-rates}). Pairwise agreement ranges from $94.1\%$ to $97.1\%$, with $\kappa$ from $0.774$ to $0.877$ (Table~\ref{tab:judge-agreement}). A repeated Gemini~2.5~Pro run agrees with itself on $94.6\%$ of parseable cases, providing context for the remaining cross-judge variation.

These results support patch anchoring as the criterion aligned with the corpus's intended label semantics, rather than as a relaxation introduced to increase acceptance. We therefore apply it to all $4{,}900$ candidates using Gemini~2.5~Flash as the production judge for throughput. Its $82.9\%$ acceptance on the audit sample is close to the $83.9\%$ first-pass acceptance observed at corpus scale. We report $\kappa$ as an agreement statistic rather than as a direct measure of judge quality because the original human judgments used a different interpretation of the task and the patch-anchored labels are imbalanced.

\begin{table}[tbp]
  \centering
  \caption{Acceptance rate on the $170$-pair audit sample, by rubric.}
  \label{tab:rubric-rates}
  \small
  \begin{tabular}{llr}
    \toprule
    Rubric & Source & Accepted \\
    \midrule
    \multirow{4}{*}{Whole-method}
      & Annotator~B (human)      & 28.2\% \\
      & GPT-5.5                  & 32.4\% \\
      & Annotator~A (human)      & 57.1\% \\
      & Gemini~2.5~Flash         & 64.7\% \\
    \addlinespace
    \midrule
    \addlinespace
    \multirow{3}{*}{Patch-anchored}
      & Gemini~2.5~Flash         & 82.9\% \\
      & Gemini~2.5~Pro           & 85.9\% \\
      & GPT-5.5                  & 86.5\% \\
    \bottomrule
  \end{tabular}
\end{table}

\begin{table}[tbp]
  \centering
  \caption{Judge-verdict agreement under the patch-anchored rubric.}
  \label{tab:judge-agreement}
  \small
  \begin{tabular}{llrrr}
    \toprule
    Judge pair & Sample & $n$ & $\kappa$ & Agree \\
    \midrule
    Gemini~2.5~Pro vs.\ GPT-5.5    & audit sample             & 170 & 0.877 & 97.1\% \\
    Gemini~2.5~Pro vs.\ GPT-5.5    & audit sample, excl.\ ISE & 150 & 0.916 & 98.7\% \\
    Gemini~2.5~Flash vs.\ Pro      & audit sample             & 170 & 0.799 & 94.7\% \\
    Gemini~2.5~Flash vs.\ Pro      & audit sample, excl.\ ISE & 150 & 0.843 & 97.3\% \\
    Gemini~2.5~Flash vs.\ GPT-5.5  & audit sample             & 170 & 0.774 & 94.1\% \\
    Gemini~2.5~Flash vs.\ Pro      & SIOOBE, full type      & 498$^{a}$ & 0.898 & 96.8\% \\
    Gemini~2.5~Pro vs.\ itself     & audit sample, repeated & 167$^{a}$ & n/a   & 94.6\% \\
    \bottomrule
    \multicolumn{5}{l}{\footnotesize $^{a}$Judgements returning unparseable
      output are excluded ($2$ of $500$ \SIOOBE; $3$ of $170$ repeats).} \\
  \end{tabular}
\end{table}

\subsection{Full-Corpus Validation and Repair}
\label{sec:val:final}
Applying the patch-anchored rubric to all $4{,}900$ generated candidates accepts $4{,}112$ pairs on the first pass, or $83.9\%$ (Table~\ref{tab:validation-composition}). Rejections fall into two categories. If Stage~1 finds that the buggy method does not throw the labelled exception at the patched site, the pair is rejected permanently because it does not instantiate the claimed defect. If Stage~1 succeeds but Stage~2 finds that the repair does not remove the identified risk, the defect is present but the generated fix is inadequate, so the pair is eligible for one repair attempt.

Only fix-side failures enter the repair loop. Claude Opus~4.7 regenerates the fix once using the production judge's identified throw site, trigger, and rejection reason; the buggy method is unchanged. The resulting pair is judged once under the same patch-anchored criterion. This procedure was fixed in advance: there is no iterative repair, no second regeneration attempt, and no repair for pairs that fail the buggy-side check. Of $168$ repair candidates, $163$ pass on re-judgment. The final released corpus therefore contains $4{,}275$ pairs, for an overall acceptance rate of $87.2\%$.

\ISE is the least reliable exception type under both production judging and the cross-judge analysis of Section~\ref{sec:val:robustness}. Only $305$ of its $500$ candidates pass initially, and $100$ of the final $405$ released pairs enter through the predefined same-judge repair-and-readmit procedure. Because \ISE is often raised deliberately to enforce object-state preconditions, distinguishing unintended from intentional throws is also difficult. We therefore flag all $405$ released \ISE pairs so that studies requiring more uniform validation confidence can exclude them. Section~\ref{sec:availability} gives the corresponding dataset-use guidance.

The remaining $625$ candidates are released separately with their judge verdicts and rejection reasons so that the exclusion decisions can be audited and the rejected population can be studied.

\begin{table}[tbp] 
  \centering
  \caption{Validation and repair outcome per exception type.}
  \label{tab:validation-composition}
  \small
  \setlength{\tabcolsep}{4pt}
  \begin{tabular}{lrrrrrrrr}
    \toprule
    & & \multicolumn{2}{c}{Initial valid} & \multicolumn{2}{c}{Repair} & &
    \multicolumn{2}{c}{Final} \\
    \cmidrule(lr){3-4}\cmidrule(lr){5-6}\cmidrule(lr){8-9}
    Exception type & Judged & $n$ & \% & cand. & acc. & Excl. & $n$ & \% \\
    \midrule
    \texttt{NullPointerException}            & 1{,}000 & 975 & 97.5 &   1 &   1 &  24 &   976 & 97.6 \\
    \texttt{IllegalArgumentException}        &   800 & 649 & 81.1 &  21 &  20 & 131 &   669 & 83.6 \\
    \texttt{ClassCastException}              &   700 & 646 & 92.3 &   8 &   8 &  46 &   654 & 93.4 \\
    \texttt{NumberFormatException}           &   700 & 568 & 81.1 &  28 &  28 & 104 &   596 & 85.1 \\
    \texttt{ArrayIndexOutOfBoundsException}  &   700 & 570 & 81.4 &   5 &   5 & 125 &   575 & 82.1 \\
    \texttt{IllegalStateException}           &   500 & 305 & 61.0 & 104 & 100 &  95 &   405 & 81.0 \\
    \texttt{StringIndexOutOfBoundsException} &   500 & 399 & 79.8 &   1 &   1 & 100 &   400 & 80.0 \\
    \midrule
    Total                                    & 4{,}900 & 4{,}112 & 83.9 & 168 & 163 & 625 & 4{,}275 & 87.2 \\
    \bottomrule
  \end{tabular}
\end{table}
\subsection{Independent JVM Verification}
\label{sec:val:execution}
Judgment-based validation establishes whether a pair satisfies the patch-anchored criterion, but it does not show that the labelled failure can be reproduced under a real JVM. We therefore add an execution check. A pair is confirmed when a concrete input makes the buggy method raise the labelled exception at the patched site and the repaired method no longer raises it there. The semantic and execution checks therefore use the same patch-local criterion.

\runin{Harness construction.}
We evaluate three harness-construction regimes (Table~\ref{tab:exec-runs}). The judge-fed regime receives both methods together with the production judge's recorded site, trigger, and explanation. It tests whether the judge's account can be reproduced, but is not independent of that account. Our primary regime instead uses GPT-5.5, from a different model family than both the Gemini production judge and the Claude corpus generator. It receives only the two methods and their patch diff, not the judge's site, trigger, or explanation, and must derive its own triggering input. A third methods-only regime also removes the patch diff and serves as a localization control. All generated harnesses preserve the method bodies verbatim, add only the stubs needed for execution, and run under a stock JDK~11.

The independent builders are not allowed to manufacture the labelled failure by forcing a stubbed collaborator to return an exceptional value such as null. If no valid trigger can be constructed under this constraint, the builder declines the case. We count these cases as unconfirmed rather than as evidence that the pair is invalid.

\runin{Independent execution result.}
On a stratified sample of $500$ released pairs, the diff-aided independent builder confirms $421$ pairs, or $84.2\%$ (Wilson $95\%$ CI $[80.7,87.1]$; Tables~\ref{tab:exec-runs} and~\ref{tab:exec-pertype}). Sixty cases receive no usable trigger; in $44$ of them, confirmation would require forcing a stubbed collaborator into an exceptional state prohibited by the independence constraint. We therefore interpret $84.2\%$ as a conservative confirmation rate, not as evidence that the remaining $15.8\%$ are invalid.

Confirmation varies by exception type (Table~\ref{tab:exec-pertype}), from $100\%$ for \SIOOBE to $53.2\%$ for \ISE. The lower rates for \IAE and \ISE occur in types where failure often depends on state-validation logic behind collaborators that the independent harness is not allowed to manipulate artificially. These execution limits do not replace the corpus-wide semantic validation of Sections~\ref{sec:val:semantic}--\ref{sec:val:final}.

\runin{Why the diff-aided arm is primary.}
The regime comparison shows two competing risks. Giving the builder the production judge's account produces more harness-induced and off-method failures, consistent with the harness being steered toward that account. Removing the judge's account avoids these artifacts. Removing the patch diff as well, however, increases cases where the harness reaches a different occurrence of the same exception rather than the site addressed by the patch. The diff-aided regime therefore provides the best balance: it remains independent of the semantic judge while retaining enough information to localise the defect addressed by the patch.

We therefore use the diff-aided independent arm as the primary JVM check. The blinded human audit motivated the patch-anchored criterion, while the cross-judge comparison and JVM execution test that criterion through different sources of evidence.
\begin{table}[t]
\centering
\small
\caption{Execution-based validation under three harness-construction regimes, with the sample size reported separately for each regime. Criteria: (a) the buggy method raises the labelled exception; (b) (a) holds and the fixed method raises nothing; (c) the buggy method raises it at the patched site and the fixed method does not.}
\label{tab:exec-runs}
\begin{tabular}{lllrrrr}
\toprule
Regime & Builder & Builder input & $n$ & (a) buggy-side & (b) both-sides & (c) site-anchored \\
\midrule
Judge-fed    & Claude Opus~4.7 & methods, site, trigger, explanation & 151 & 148/151 (98.0\%) & 129/151 (85.4\%) & 126/151 (83.4\%) \\
Diff-aided   & GPT-5.5         & methods, patch diff                 & 500 & 443/500 (88.6\%) & 401/500 (80.2\%) & 421/500 (84.2\%) \\
Methods-only & GPT-5.5         & methods                             & 129 & 119/129 (92.2\%) & 106/129 (82.2\%) & 110/129 (85.3\%) \\
\bottomrule
\end{tabular}
\end{table}

\begin{table}[t]
\centering
\small
\caption{Per-type site-anchored confirmation (criterion (c)) for the diff-aided independent arm ($n=500$), with Wilson $95\%$ confidence intervals.}
\label{tab:exec-pertype}
\begin{tabular}{lrrr}
\toprule
Exception type & $n$ & (c) confirmed & Wilson 95\% CI \\
\midrule
StringIndexOutOfBoundsException & 47  & 47/47 (100.0\%) & [92.4, 100.0] \\
NumberFormatException           & 70  & 68/70 (97.1\%)  & [90.2, 99.2] \\
ArrayIndexOutOfBoundsException  & 68  & 66/68 (97.1\%)  & [89.9, 99.2] \\
ClassCastException              & 77  & 70/77 (90.9\%)  & [82.4, 95.5] \\
NullPointerException            & 114 & 93/114 (81.6\%) & [73.5, 87.6] \\
IllegalArgumentException        & 77  & 52/77 (67.5\%)  & [56.5, 76.9] \\
IllegalStateException           & 47  & 25/47 (53.2\%)  & [39.2, 66.7] \\
\midrule
Overall                         & 500 & 421/500 (84.2\%) & [80.7, 87.1] \\
\bottomrule
\end{tabular}
\end{table}

\runin{Rejected audit cases.}
The execution check applies only to released pairs and therefore cannot test Stage~1 audit cases for which no patch-local trigger was recorded. Both human annotators had independently scored all $19$ rejected audit pairs; each judged $18$ of the $19$ invalid, with their disagreements occurring on different pairs. This agreement provides an additional check on the exclusion decisions. The detailed rejection taxonomy and judge records are included in the replication package.

\begin{finding}{1}
\emph{Under the patch-anchored criterion, $83.9\%$ of the $4{,}900$ candidates pass on first judgment and $87.2\%$ after one predefined fix-only repair. Pairwise agreement among three judges from two model families ranges from $94.1\%$ to $97.1\%$, and an independently generated JVM harness confirms $84.2\%$ of a $500$-pair stratified sample. \ISE remains the least reliable exception type and is explicitly flagged in the release.}
\end{finding}

\section{Corpus Fidelity (RQ2)}
\label{sec:fidelity}
\label{sec:distribution}
Validity tells us whether a pair represents the exception it is labelled with. It does not tell us whether the generated methods resemble the real bugs they are intended to extend. This distinction matters because distribution mismatch can hurt transfer~\cite{alrashedy2024learning,ren2025few}. We therefore compare RXSynth with two real reference pools, first structurally and then lexically, and identify where the match breaks down.

\subsection{Methodology}
\label{sec:dist:method}
We compare the released corpus with the real bug/fix pairs using seven structural metrics: physical lines of code (PLOC; non-blank lines), logical lines of code (LLOC; statement nodes), extended cyclomatic complexity, maximum nesting depth, identifier occurrences, formal parameters, and lexical tokens. We compute each metric identically for the synthetic and real methods. Identifier and token counts are occurrences rather than distinct vocabulary; Section~\ref{sec:dist:token} examines lexical composition separately. We measure buggy and fixed methods separately and, unless stated otherwise, report the buggy methods because they are the population a detector must characterize. When one size measure is needed, we use token count because it is formatting-invariant and directly relevant to transformer inputs.

For each metric, we report the two-sample Kolmogorov--Smirnov statistic $D$, its two-sided $p$-value, and Cliff's $\delta$~\cite{cliff1993dominance}. We interpret $|\delta|$ using Romano's thresholds~\cite{romano2006appropriate}: negligible below $0.147$, small below $0.330$, medium below $0.474$, and large otherwise. We treat effect sizes as the primary evidence. The synthetic sample ($n=4{,}275$) is much larger than either real reference, and generated methods are clustered by their conditioning seeds. We therefore treat $p$-values as descriptive and do not interpret non-significance, or a negligible $\delta$, as proof of equivalence.

We use two real reference sets. The \emph{seed pool} contains all $170$ extracted real pairs, whereas the \emph{157-pair fidelity reference set} contains the $157$ pairs remaining after the upper-tail reference exclusions of Section~\ref{sec:seeds:generation-pool}. Comparison with the fidelity reference set measures fidelity to a fixed real reference population drawn from the intended conditioning range; as Section~\ref{sec:seeds:generation-pool} records, it is not an exact reconstruction of the exemplars that actually conditioned production generation. Comparison with all $170$ seeds adds the $13$ upper-tail reference exclusions and shows how fidelity estimates change when those larger real methods are included. We report both throughout this section, but do not treat their difference as an additive decomposition of generator error.

\subsection{Structural Distribution}
\label{sec:dist:structural}

\runin{Approach.}
We compute the seven structural metrics over all $4{,}275$ released buggy methods and over each real reference set, first pooled across exception types and then within type.

\begin{figure}[tbp]
  \centering
  \includegraphics[width=\linewidth]{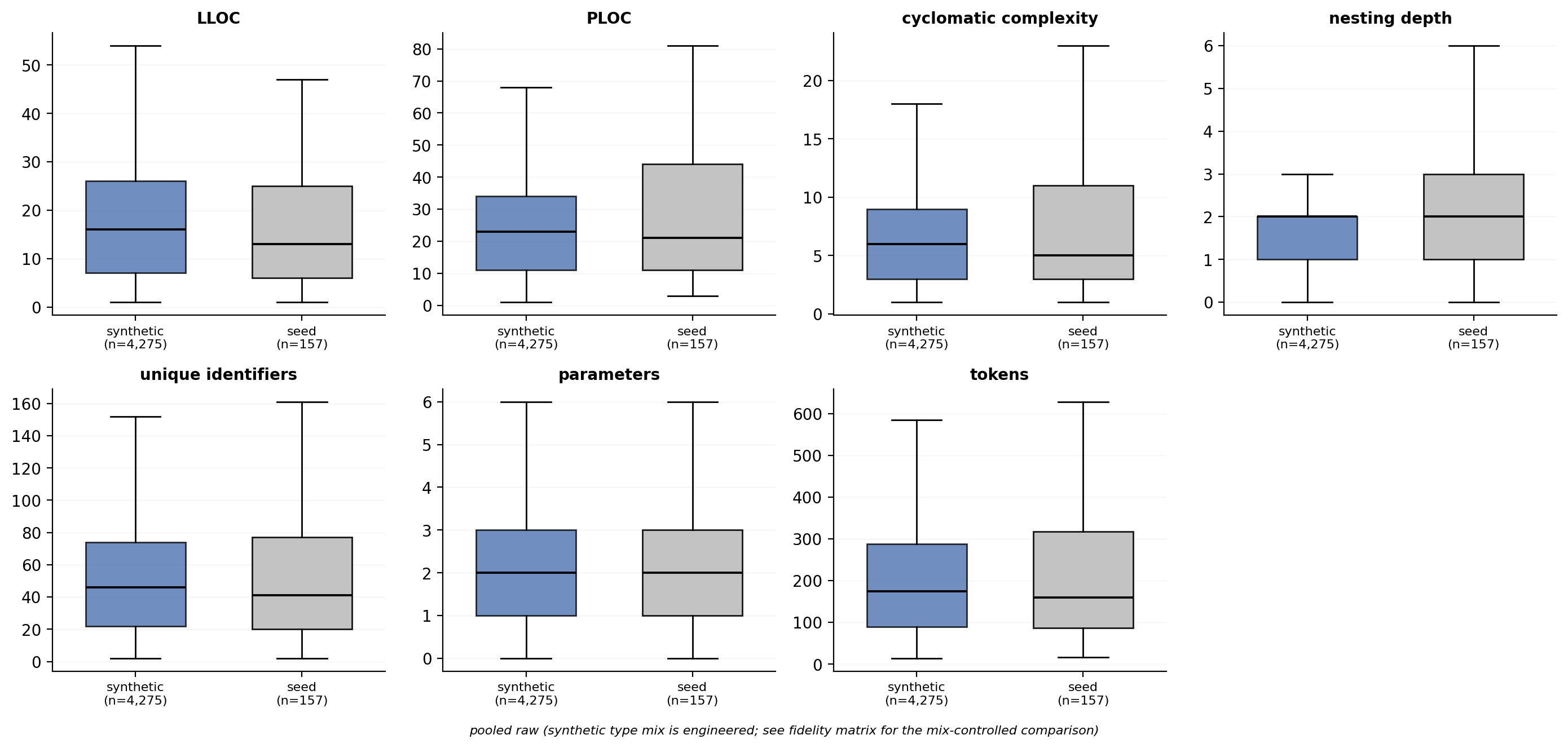}
  \caption{Pooled distributions of the seven structural metrics for the released corpus against the $157$-pair fidelity reference set.}
  \label{fig:pooled-metrics}
\end{figure}

\runin{Pooled result.}
RXSynth is structurally close to the $157$-pair fidelity reference set (Table~\ref{tab:fidelity-pooled} and Figure~\ref{fig:pooled-metrics}). All seven pooled Cliff's $\delta$ values are negligible, and five are below $0.02$ in absolute value. Parameter count is the only metric with a statistically detectable shift ($D=0.160$, $p<0.001$, $\delta=+0.09$). Because the effect size remains negligible under our prespecified threshold, we treat this as a small but systematic artifact rather than a large mismatch. This is consistent with generated methods more often exposing required state as explicit parameters.

The pooled comparison also reflects corpus composition. RXSynth is deliberately more balanced than the \NPE-dominated real pool, so pooled medians can differ even when within-type distributions are similar. We therefore examine each exception type separately below.

Against the complete $170$-seed pool, the match becomes weaker. All seven KS tests are significant, and $D$ increases most strongly for identifier occurrences, cyclomatic complexity, tokens, and nesting, by $72$--$95\%$ relative to the fidelity-reference-set comparison. The two reference sets differ by the $13$ upper-tail reference exclusions of Section~\ref{sec:seeds:generation-pool}, so the larger distances show sensitivity to including those larger real methods. They should not be interpreted as the effect of completely withholding those pairs from production generation, because two \texttt{JacksonCore-19} reference exclusions did condition released generations. We do not treat the change in $D$ as an additive decomposition of generator error; Section~\ref{sec:preferred-band} examines the size-dependent departure directly.

\begin{figure}[tbp]
  \centering
  \includegraphics[width=\linewidth]{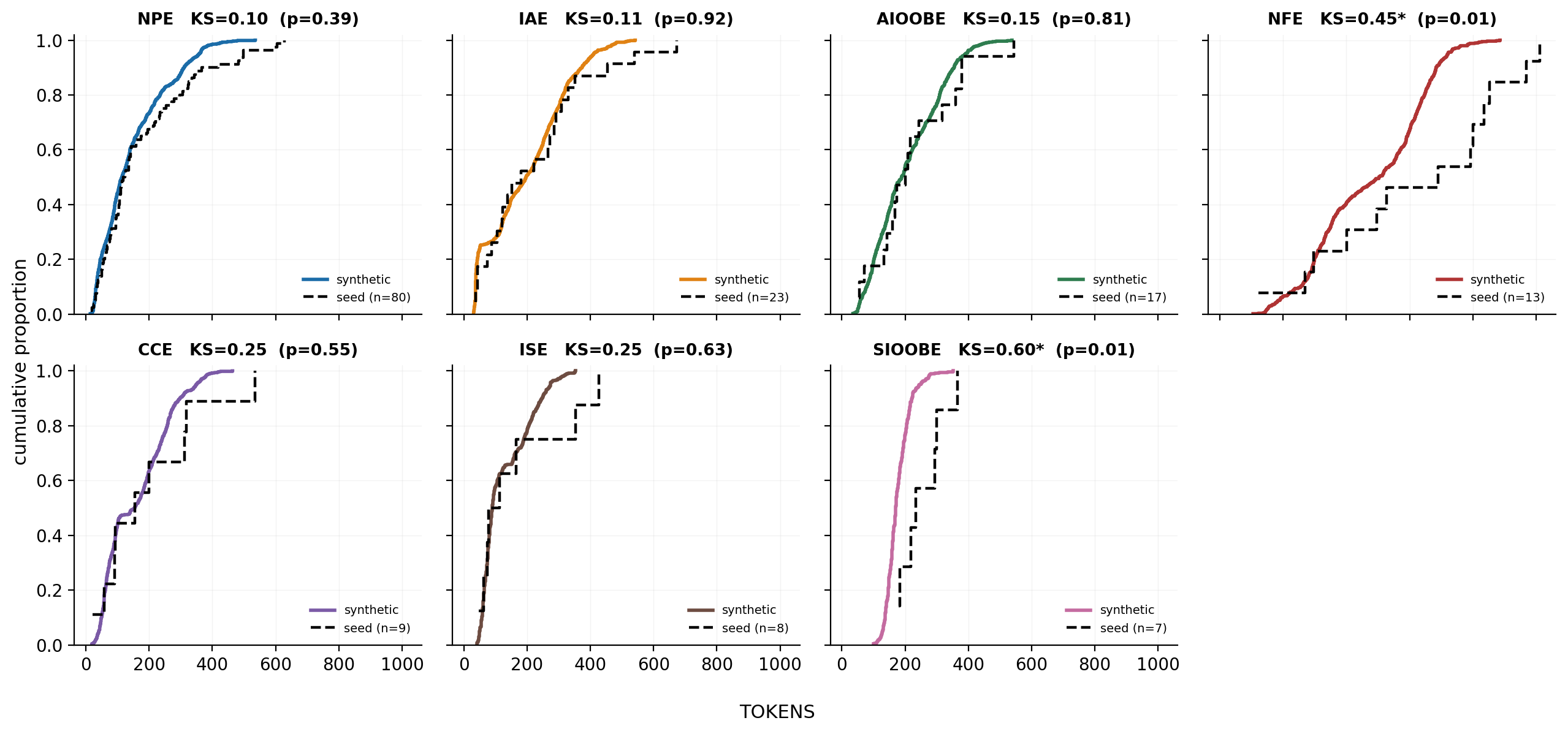}
  \caption{Per-type token-count ECDFs of the released corpus against the $157$-pair fidelity reference set.}
  \label{fig:per-type-ecdf}
\end{figure}

\runin{Per-type departures.}
Five of the seven exception types show no statistically detectable token-count difference from the corresponding seeds in the $157$-pair fidelity reference set, although the real samples are too small to establish equivalence (Figure~\ref{fig:per-type-ecdf}). The two exceptions are \NFE and \SIOOBE, and both are smaller than their real reference methods.

For \NFE, token count differs with $D=0.45$ and $\delta=-0.33$ (median $503$ versus $689$), and LLOC also decreases ($D=0.43$). This indicates genuine statement compression. \SIOOBE shows a larger token difference ($D=0.60$, $\delta=-0.72$; median $170$ versus $233$), but its LLOC distribution is not detectably different. Instead, it also differs in PLOC and identifier occurrences. Its generated methods therefore contain roughly as many statements as the seeds but less incidental work within those statements.

The two types reveal different fidelity limits. \NFE mainly shows method-size compression, whereas \SIOOBE shows reduced code density and surrounding context. Section~\ref{sec:preferred-band} returns to both patterns when we examine fidelity as a function of exemplar size.

\begin{table}[tbp]
  \centering
  \caption{Structural comparison of the released corpus against two reference
  sets, pooled over exception types.}
  \label{tab:fidelity-pooled}
  \small
  \begin{tabular}{lrrrrrrr}
    \toprule
    & \multicolumn{1}{c}{Corpus}
    & \multicolumn{3}{c}{vs.\ fidelity reference set ($n{=}157$)}
    & \multicolumn{3}{c}{vs.\ seed pool ($n{=}170$)} \\
    \cmidrule(lr){2-2}\cmidrule(lr){3-5}\cmidrule(lr){6-8}
    Metric & med & med & $D$ & $\delta$ & med & $D$ & $\delta$ \\
    \midrule
    Identifier occurrences &  46 &  41 & 0.062       & $+0.01$ & 45.5 & 0.121$^{*}$ & $-0.06$ \\
    Cyclomatic complexity  &   6 &   5 & 0.066       & $+0.00$ &    6 & 0.123$^{*}$ & $-0.07$ \\
    Tokens                 & 174 & 160 & 0.067       & $-0.01$ &  182 & 0.119$^{*}$ & $-0.08$ \\
    Nesting depth          &   2 &   2 & 0.071       & $-0.02$ &    2 & 0.122$^{*}$ & $-0.08$ \\
    PLOC                   &  23 &  20 & 0.101       & $-0.01$ & 23.5 & 0.158$^{*}$ & $-0.08$ \\
    LLOC                   &  16 &  13 & 0.108       & $+0.07$ & 13.5 & 0.113$^{*}$ & $-0.01$ \\
    Parameters             &   2 &   2 & 0.160$^{*}$ & $+0.09$ &    2 & 0.167$^{*}$ & $+0.09$ \\
    \bottomrule
    \multicolumn{8}{l}{\footnotesize $^{*}$ two-sided KS $p < 0.05$;
      $n = 4{,}275$ synthetic buggy methods.} \\
  \end{tabular}
\end{table}

\subsection{Lexical Distribution}
\label{sec:dist:token}
Section~\ref{sec:dist:structural} compares method size and shape. We next ask whether generated methods use the same exception-relevant Java constructs as the $157$-pair fidelity reference set. Direct vocabulary comparison would be dominated by project-specific identifiers that synthetic methods cannot reproduce by construction. We therefore count thirteen semantic constructs: null literals, equality and inequality checks, \texttt{instanceof}, casts, array accesses, \texttt{try} and \texttt{catch}, explicit \texttt{throw}, \texttt{return null}, \texttt{equals} calls, division, and modulo. We count each construct per buggy method and compare the distributions within exception type, yielding $85$ non-empty type-by-construct cells. We use the Mann--Whitney $U$ test with Benjamini--Hochberg false-discovery-rate control at $q=0.05$~\cite{benjamini1995controlling}.

\runin{Result.}
Seventy-four of the $85$ cells show no statistically detectable difference. Eleven differ after FDR correction, and all eleven move in the same direction: the synthetic corpus contains fewer occurrences than the real methods (Table~\ref{tab:markers}). No construct is significantly over-produced in any exception type.

The significant differences are concentrated in constructs incidental to the labelled exception rather than in those that directly realise it. For example, casts, array accesses, \texttt{instanceof}, and \texttt{try}/\texttt{catch} constructs are under-produced when they form surrounding program context rather than the failure mechanism itself. Table~\ref{tab:markers} gives the affected type-by-construct cells. By contrast, we detect no significant divergence for the defining constructs of the target exceptions: null-related constructs for \NPE, array accesses for \AIOOBE, casts for \CCE, or explicit \texttt{throw} statements for \IAE and \ISE. Because the real per-type samples are small, these non-significant results should not be interpreted as evidence of equivalence.

This pattern suggests that grounding preserves the defect mechanism better than the incidental work surrounding it. Synthetic methods reproduce the constructs that realise the labelled failure, but more often omit unrelated operations present in production methods. This complements the structural result of Section~\ref{sec:dist:structural}: \SIOOBE in particular preserves statement count while carrying fewer tokens and fewer incidental constructs per statement. Section~\ref{sec:preferred-band} examines this density effect directly across exemplar sizes.

\begin{table}[tbp]
  \centering
  \caption{Language constructs significantly under-produced in the corpus
  relative to the $157$-pair fidelity reference set (Benjamini--Hochberg, $q < 0.05$).}
  \label{tab:markers}
  \small
  \begin{tabular}{@{}l r r r@{\hspace{2em}} l r r r@{}}
    \toprule
    Construct & Corpus & Real & Ratio & Construct & Corpus & Real & Ratio \\
    \midrule
    \multicolumn{4}{@{}l}{\textit{\NPE}} & \multicolumn{4}{@{}l}{\textit{\ISE}} \\
    casts               & 0.07 & 0.46 & 0.15 & casts               & 0.02 & 0.25 & 0.08 \\
    array access        & 0.06 & 0.35 & 0.17 & \texttt{instanceof} & 0.00 & 0.12 & 0.00 \\
    \texttt{catch}      & 0.03 & 0.14 & 0.21 &                     &      &      &      \\
    \texttt{instanceof} & 0.01 & 0.08 & 0.13 &                     &      &      &      \\
    \midrule
    \multicolumn{4}{@{}l}{\textit{\AIOOBE}} & \multicolumn{4}{@{}l}{\textit{\SIOOBE}} \\
    \texttt{instanceof} & 0.05 & 0.35 & 0.14 & array access & 0.02 & 0.86 & 0.02 \\
    \texttt{try}        & 0.00 & 0.06 & 0.00 & casts        & 0.01 & 0.14 & 0.07 \\
    \texttt{catch}      & 0.00 & 0.06 & 0.00 &              &      &      &      \\
    \bottomrule
    \multicolumn{8}{@{}l}{\footnotesize Mean occurrences per method; ratio is
      corpus over real.} \\
  \end{tabular}
\end{table}

\runin{Repair style.}
We also test whether generated repairs resemble real repairs in their use of explicit guards. A fix is counted when it introduces a \texttt{throw} statement absent from its buggy counterpart. The rates are nearly identical: $295$ of $4{,}275$ synthetic fixes ($6.9\%$) introduce at least one new throw, compared with $12$ of $170$ real fixes ($7.1\%$, Wilson $95\%$ CI $[4.1,11.9]$)~\cite{wilson1927probable}. \IAE dominates this repair idiom in both populations, accounting for $309$ of $373$ introduced throws in the corpus and $9$ of $15$ in the seeds. Thus, beyond buggy-side lexical composition, the corpus also reproduces the overall frequency of repairs that add explicit validation throws.

\subsection{Fixed-Generator Grounding Ablation}
\label{sec:dist:confound}

The production zero-shot and seed-grounded campaigns differ in both conditioning and generating model, so their structural difference alone cannot identify which factor improves fidelity. We therefore run a separate fixed-generator ablation in which GPT-4o generates both arms. This experiment is not part of the released corpus; its purpose is to compare the seed-grounded and unconditioned generation configurations while holding model choice constant.

\runin{Approach.}
We compare three \NPE buggy-method LLOC distributions: $1{,}000$ unconditioned GPT-4o generations, $380$ GPT-4o generations produced under the seed-grounded configuration, and the $87$ real \NPE seed methods. GPT-4o was not the generator selected for corpus production, so the comparison does not reuse the model whose alignment motivated the final campaign.

\begin{figure}[tbp]
  \centering
  \includegraphics[width=0.6\linewidth]{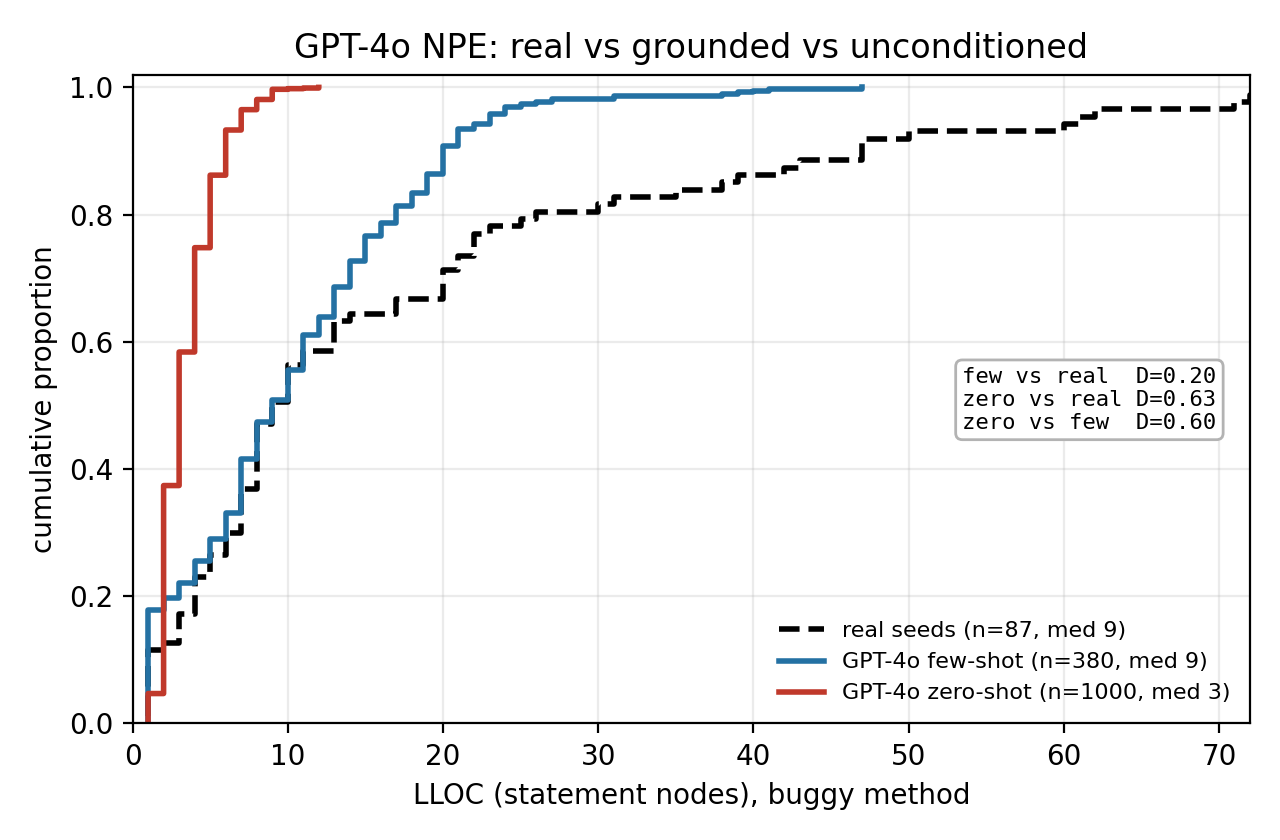}
  \caption{Buggy-method size under a fixed GPT-4o generator for the unconditioned and seed-grounded configurations, compared with the $87$ real \NPE seeds.}
  \label{fig:grounding-confound}
\end{figure}

\runin{Result.}
Without grounding, GPT-4o collapses toward short methods: the median is $3$ statements, compared with $9$ for the real seeds, and the Kolmogorov--Smirnov distance to the real distribution is $D=0.63$ (Figure~\ref{fig:grounding-confound}). Under the seed-grounded configuration, the median rises to $9$ statements and the distance falls to $D=0.20$. The two generated distributions themselves differ at $D=0.60$ with Cliff's $\delta=-0.54$, a large effect.

Because the generator is identical in both arms, the observed difference cannot be explained by generator selection. The seed-grounded arm supplies real exemplars together with exemplar-derived size targets, so the ablation tests this configuration as a whole. It does not separate the contribution of exemplar content from that of size conditioning. We therefore treat the result as evidence that the seed-grounded configuration yields better structural alignment for \NPE under GPT-4o, not as a mechanism test for either component or for all exception types.

The seed-grounded configuration does not eliminate the mismatch entirely. Although it recovers the center of the real distribution, the grounded output still under-represents its upper tail, leaving $D=0.20$ from the real seeds. Section~\ref{sec:preferred-band} examines this residual directly and shows where exemplar fidelity begins to degrade with method size.

\subsection{Distance from Seeds}
\label{sec:dist:seed-distance}

Because generation is conditioned on real bug-fix exemplars, we check whether synthetic methods reproduce those exemplars directly. For every buggy and fixed synthetic method, we compute Jaccard similarity to the nearest real seed method of the same exception type under two representations: unordered token sets, which capture shared vocabulary, and contiguous token 5-grams, which are more sensitive to copied code sequences. Of the resulting $8{,}550$ nearest-neighbour comparisons, $94$ ($1.10\%$) differ depending on whether the reference is the $157$-pair fidelity reference set or the full $170$-pair seed pool, so a small number of synthetic methods do have a reference-excluded seed as their nearest neighbour. The differences are small in magnitude and do not change the conclusions below; we report the fidelity-reference-set values.

We observe no exact reproductions. The maximum token 5-gram Jaccard similarity is $0.448$; thus the closest observed synthetic--seed pair shares less than half of its contiguous token 5-grams under this measure. Shared token-set vocabulary is expected within an exception type because null checks, casts, bounds guards, and related idioms recur across implementations.

This result addresses instance-level copying only. Synthetic methods may still inherit structural and lexical characteristics from the seeds that conditioned them. Section~\ref{sec:leave-seeds-out} therefore controls provenance explicitly when evaluating transfer to real defects.

\subsection{Fidelity to the Conditioning Exemplar}
\label{sec:preferred-band}

Sections~\ref{sec:dist:structural} and~\ref{sec:dist:token} identify two residual mismatches: large methods are under-represented, and generated methods contain less incidental work than their real counterparts. Each synthetic pair retains source-bug provenance and exemplar-location metadata for the examples that conditioned it, allowing these effects to be tested at exemplar level rather than only through aggregate corpus distributions. Exact method-pair attribution is ambiguous for \texttt{Lang-47}: its two real method pairs have different source spans but the same recorded exemplar size (13 LLOC and 18 PLOC). Since the present analysis uses exemplar size, that ambiguity does not affect the size-ratio comparison.

\runin{Approach.}
For each of the $4{,}275$ released pairs, we compute generated method size relative to the mean size of its conditioning exemplars and group pairs by exemplar size. A ratio near one indicates that generation preserves exemplar size. We measure both logical lines of code (LLOC), which counts statements, and physical lines of code (PLOC), which also reflects how densely those statements are expressed.

\begin{figure}[tbp]
  \centering
  \includegraphics[width=0.5\linewidth]{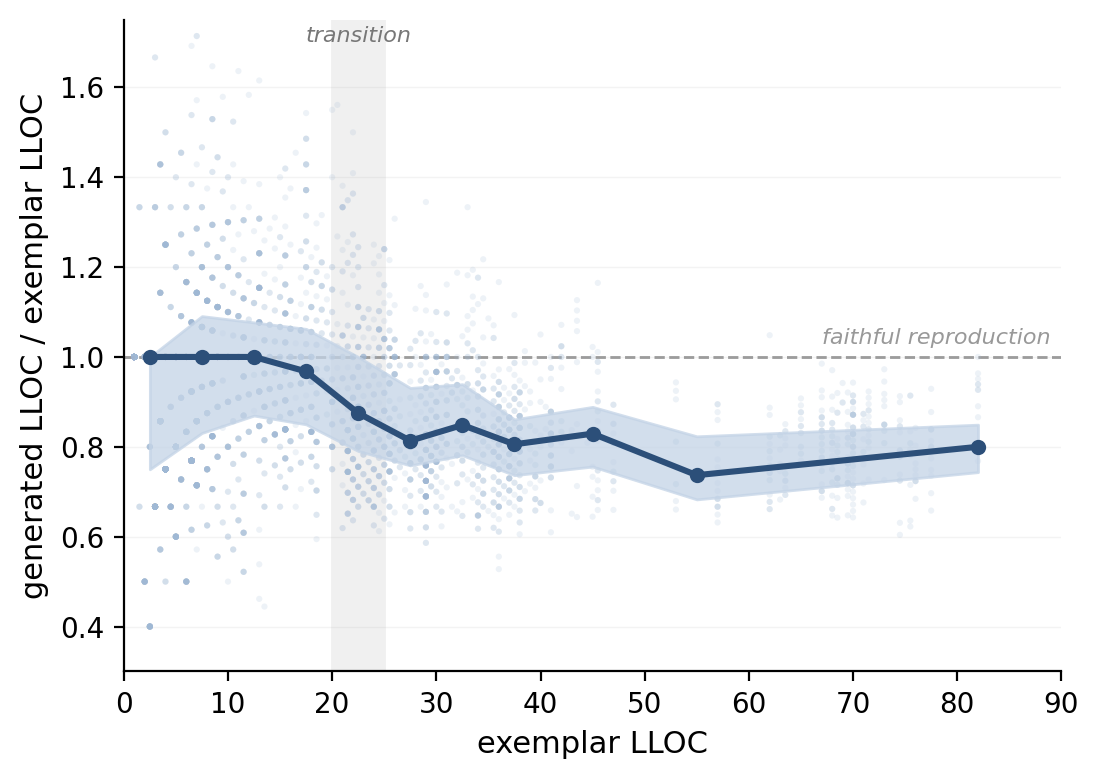}
  \caption{Generated statement count relative to the conditioning exemplar. Points are individual pairs, the line is the binned median, and the shaded band is the interquartile range; the vertical axis is cropped.}
  \label{fig:compression}
\end{figure}

\runin{Two fidelity regimes.}
The LLOC profile shows two regimes (Figure~\ref{fig:compression}). For exemplars below $15$ statements, the median generated-to-exemplar ratio is $1.00$; between $15$ and $20$ statements it is $0.97$. Across these smaller methods, generation therefore reproduces statement count closely. Above roughly $20$ statements, fidelity changes. For exemplars above $25$ statements, the median ratio stabilises around $0.81$, ranging from $0.74$ to $0.85$ across size bands. Generated methods continue to grow with their exemplars, from a median of $5$ statements for the smallest exemplars to $56$ for the largest, so this is not a hard output-length ceiling. It is proportional compression: the generator continues to track larger exemplars while retaining roughly four fifths of their statement structure.

PLOC reveals a second, distinct effect. Synthetic methods use a median of $1.43$ physical lines per statement, compared with $1.75$ for the $157$-pair fidelity reference set, and the direction is consistent across all seven exception types. The generator therefore expresses comparable statement-level logic more densely even where statement count itself is preserved. This density difference is present across method sizes, whereas statement loss emerges primarily once exemplars move beyond the roughly $20$-statement range.

The two effects also explain the type-specific departures observed earlier. \NFE provides the clearest example of statement compression: its conditioning methods are large, with a median of $57$ statements, and none of its seeds is among the $13$ upper-tail reference exclusions. Its smaller generated methods therefore reflect generator behaviour rather than the reference-set upper-tail distinction. \SIOOBE shows the complementary case. Its exemplars have a median of $19$ statements, near the boundary at which statement loss begins, so their statement counts are largely preserved while the generated methods remain substantially denser. This matches the lexical result of Section~\ref{sec:dist:token}: the missing content is primarily incidental work surrounding the defect rather than the exception mechanism itself.

The same profile supports treating the upper tail separately in the fidelity analysis. The $13$ upper-tail reference exclusions have a median size of $67$ statements, compared with $13$ for the $157$ pairs in the fidelity reference set, placing them well within the regime where generation compresses exemplar structure. This is a statement about the reference-set design, not complete removal from production generation: two \texttt{JacksonCore-19} reference exclusions did condition released generations. The full-$170$ comparison therefore complements the within-range $157$-pair reference by retaining the influence of those larger methods.

Taken together, the results place a concrete boundary on seed-grounded synthesis. Within the represented size range, grounding closely reproduces the statement structure of its exemplars. Beyond that range, it continues to follow exemplar size but does so proportionally rather than faithfully, while expressing code more densely at all sizes. The observed fidelity supports close interpolation within the structural range the generator can reproduce, but does not by itself make the generator extrapolate faithfully to substantially larger or more context-rich methods.

\begin{finding}{2}
\emph{RXSynth is structurally close to the $157$-pair fidelity reference set within the represented size range, and pair-level analysis shows close size tracking for smaller recorded exemplars. Fidelity weakens for larger methods: generated methods increasingly compress exemplar structure above roughly $20$ statements and contain less incidental surrounding work even when statement count is preserved. The observed fidelity supports close interpolation within the represented range, but does not guarantee faithful extrapolation to substantially larger or more context-rich methods.}
\end{finding}

\section{Empirical Study: Runtime-Exception Prediction Under Two Information Settings}
\label{sec:usecase}

RXSynth lets us study what information is available for runtime-exception prediction at a scale the $170$ real pairs alone cannot support. We evaluate two information settings on the same real pairs, applying the same source-bug-level provenance-exclusion principle despite differences in fold construction. RQ3 gives the model one method in isolation and asks whether it is defective. RQ4 gives the model both members of the pair and asks which one is defective. The pair-discrimination setting is not simply an easier version of RQ3 because it changes the information available to the model.

\subsection{Study Design}
The primary supervised experiments train on RXSynth and evaluate on the same $170$ real bug/fix pairs from Section~\ref{sec:seeds}. In the isolated-method setting (Section~\ref{sec:difficulty}), the model receives one method at a time and predicts whether it is defective or repaired. We evaluate this task with ROC-AUC over the $340$ real methods. In the pair-discrimination setting (Section~\ref{sec:contrastive}), the model receives both methods in randomised order and identifies the defective one. We evaluate this task with accuracy over the $170$ real pairs, with $50\%$ chance performance.

The two settings have different inputs, outputs, and evaluation metrics, so their ROC-AUC and pair-accuracy values should not be compared directly. Each RQ instead asks how well models perform under its own information setting. The later controls vary architecture, training source, model scale, and available information while keeping the real evaluation data fixed. Both settings use the provenance-controlled protocol described next.

\subsection{Leakage-Controlled Protocol}
\label{sec:leave-seeds-out}
\label{sec:diff:leave-seeds-out}

Seed grounding creates a provenance dependency that a random train/test split would miss. Each synthetic pair is generated from one or two real bug/fix seeds and may inherit structural, lexical, or stylistic properties of those seeds even when it is not a near-duplicate (Section~\ref{sec:dist:seed-distance}). If descendants of a real seed are used for training and that same seed is used for evaluation, performance may reflect recognition of its generation neighbourhood rather than transfer to an unseen defect. We therefore use a leave-seeds-out protocol for every synthetic-trained model.

Each synthetic record stores source-bug provenance for the real seeds used to generate it, together with additional generation metadata. The $170$ real pairs come from $161$ source bugs; eight source bugs contribute multiple method-level pairs, including one that contributes three. We assign folds at source-bug granularity, so all pairs from the same source bug remain together. The fold-construction procedures differ across the two information settings. RQ3 groups source bugs by exception type before assigning them round-robin to five folds; its archived folds are exception-stratified. In RQ4, the historical fold splitter read \texttt{error\_type} from real-seed records that instead store \texttt{exception\_canonical}. This placed all source bugs in a single stratum, producing five source-bug-grouped but not exception-stratified folds. The grouping and provenance-exclusion guarantees remain intact in both settings, but the exception mix within RQ4 folds is not controlled by stratification.

For each fold, the corresponding real pairs form the test set. Exclusion operates at source-bug granularity on both sides. Every provenance identifier recorded on a synthetic pair is mapped through a verified identifier-to-source-bug table, the held-out real pairs are mapped to the same source-bug unit, and a synthetic training pair is removed whenever its mapped provenance intersects the held-out source bugs. The mapping is table-driven rather than heuristic: identifier shapes differ across the four benchmarks, so an identifier that the table does not cover raises an error instead of passing through unmatched. No synthetic descendant of a held-out source bug therefore enters the model that evaluates that bug. Depending on the fold, $2{,}609$--$3{,}355$ synthetic training pairs remain for RQ3 and $2{,}573$--$3{,}180$ for RQ4.

Rotating over the five folds gives every real pair an out-of-fold prediction from a model trained on provenance-disjoint synthetic data. We pool these predictions for the reported real-data metrics. In RQ3, Random Forest and XGBoost use one seed-42 invocation with three internal stratified source-bug re-partitions. The neural models use three separate single-repetition runs at seeds 42, 7, and 123, each with its corresponding seeded fold assignment. RQ4 uses five run seeds (Section~\ref{sec:utility:stats}). The original runs constructed folds in memory from the run seeds. The replication package includes fold assignments subsequently regenerated from those seeds and verified against archived predictions; the original runs did not load these files.

\subsection{Isolated-Method Discrimination Remains Weak or Biased (RQ3)}
\label{sec:difficulty}
\label{sec:diff:classifier}
RQ3 asks whether the defective member of a real bug/fix pair can be identified from an isolated method body. Our main models train on RXSynth and are evaluated on the $170$ real pairs under the provenance control of Section~\ref{sec:leave-seeds-out}. A weak result from one model could reflect its architecture, the synthetic training data, or its capacity rather than the information setting. We therefore test these explanations separately.

\runin{Architecture.}
Isolated-method performance remains weak across all seven supervised classifiers. The models span feature-based methods, sequence models trained from scratch, and pre-trained code transformers; full architecture and optimisation details are provided in the replication package. Across them, pooled ROC-AUC ranges from $0.546$ to $0.610$ (Table~\ref{tab:leave-seeds-out}). Providing the queried exception type changes ROC-AUC by only $-0.008$ to $+0.020$. Neither architecture choice nor explicit access to the target exception type raises ROC-AUC above $0.610$.

\runin{Training-data source.}
Training on the real pairs does not substantially change the result. To test whether weak transfer comes from synthetic training data, we train and evaluate a Random Forest using only the real pairs. We use bug-grouped five-fold cross-validation so that methods from the same source bug never cross folds. Real-to-real ROC-AUC is approximately $0.64$, ranging from $0.636$ to $0.647$ across repeated splits (Table~\ref{tab:real-vs-synth}). This is slightly above the synthetic-trained range but remains close to it. The weak isolated-method result is therefore not specific to synthetic training data.

\runin{Model scale and inference regime.}
Zero-shot frontier models obtain higher accuracy, but their predictions remain strongly biased toward the defective class. We query three contemporary language models on each of the $340$ real methods and provide the labelled exception type (Table~\ref{tab:llm-overall}). GPT-5.5 reaches $69.4\%$ accuracy, Gemini~3.1~Pro $68.8\%$, and Llama~3.3~70B $53.5\%$. Because the evaluation contains equal numbers of defective and repaired methods, overall accuracy is also balanced accuracy. GPT-5.5 and Gemini recall almost all defective methods but correctly classify only $41.2\%$ of repaired methods as non-defective; Llama's specificity is $15.9\%$. The higher frontier-model accuracy therefore does not remove the strong defective-class bias.

\begin{table}[tbp]
  \centering
  \caption{Provenance-controlled leave-seeds-out ROC-AUC without and with the queried exception type as input (mean $\pm$ sample sd over three repetitions; protocol in Section~\ref{sec:leave-seeds-out}).}
  \label{tab:leave-seeds-out}
  \small
  \begin{tabular}{llcc}
    \toprule
    Paradigm & Model & No type & With type \\
    \midrule
    \multirow{2}{*}{Classical (features)}
      & Random Forest & $0.570 \pm 0.003$ & $0.589 \pm 0.006$ \\
      & XGBoost       & $0.575 \pm 0.004$ & $0.595 \pm 0.005$ \\
    \midrule
    \multirow{2}{*}{Sequence (from scratch)}
      & TextCNN       & $0.572 \pm 0.003$ & $0.564 \pm 0.003$ \\
      & BiLSTM        & $0.554 \pm 0.009$ & $0.558 \pm 0.017$ \\
    \midrule
    \multirow{3}{*}{Pre-trained transformer}
      & CodeBERT      & $0.597 \pm 0.017$ & $0.607 \pm 0.010$ \\
      & CodeGPT       & $0.546 \pm 0.012$ & $0.563 \pm 0.003$ \\
      & CodeT5        & $0.602 \pm 0.006$ & $0.610 \pm 0.015$ \\
    \midrule
    \multicolumn{2}{l}{Chance level} & \multicolumn{2}{c}{$0.500$} \\
    \multicolumn{2}{l}{Random Forest, provenance intact (leaky, no type)} & \multicolumn{2}{c}{$0.831$} \\
    \bottomrule
  \end{tabular}
\end{table}

\runin{Synthesis.}
The result remains limited across substantially different controls. The seven supervised architectures vary the model family, real-to-real training changes the data source, and zero-shot prompting changes both model scale and inference regime. No evaluated approach achieves strong and balanced isolated-method discrimination under the settings tested. This suggests that changing the architecture, training source, or inference regime alone does not resolve the isolated-method task in our evaluation. It does not establish that the method body contains no usable information or that reliable discrimination is impossible with other models or settings. Section~\ref{sec:contrastive} next changes the information available to the model.

\begin{table}[tbp]
\centering

\begin{minipage}[t]{0.57\textwidth}
    \centering
    \caption{Zero-shot isolated-method classification on the
    $340$ real methods ($170$ pairs), with the labelled
    exception type provided.}
    \label{tab:llm-overall}

    \small
    \setlength{\tabcolsep}{1pt}
    \renewcommand{\arraystretch}{1.40}

    \begin{tabular*}{\linewidth}{@{\extracolsep{\fill}}lccccc@{}}
        \toprule
        Model & Acc. & $F_1$ & Specificity & Precision & Recall \\
        \midrule
        GPT-5.5         & 69.4\% & 0.761 & 0.412 & 0.624 & 0.976 \\
        Gemini~3.1~Pro  & 68.8\% & 0.756 & 0.412 & 0.621 & 0.965 \\
        Llama~3.3~70B   & 53.5\% & 0.662 & 0.159 & 0.520 & 0.912 \\
        \midrule
        Always-positive & 50.0\% & 0.667 & 0.000 & 0.500 & 1.000 \\
        \bottomrule
    \end{tabular*}
\end{minipage}%
\hfill
\begin{minipage}[t]{0.40\textwidth}
    \centering
    \caption{Isolated-method ROC-AUC under three
    evaluation regimes.}
    \label{tab:real-vs-synth}

    \small
    \setlength{\tabcolsep}{1pt}
    \renewcommand{\arraystretch}{1.0}

    \begin{tabular*}{\linewidth}{@{\extracolsep{\fill}}lcc@{}}
        \toprule
        Evaluation regime & Provenance & AUC \\
        \midrule

        \shortstack[l]{Real$\to$real\\(RF, bug-grouped CV)}
            & disjoint & $0.64$ \\
        \midrule

        \shortstack[l]{Synthetic$\to$real\\(leave-seeds-out, 7 arch.)}
            & disjoint & $0.55$--$0.61$ \\
        \midrule

        \shortstack[l]{Synthetic$\to$seed\\(RF, provenance-intact)}
            & \emph{overlapping} & $0.83$ \\
        \bottomrule
    \end{tabular*}
\end{minipage}

\end{table}

\begin{figure}[tbp]
  \centering
  \includegraphics[width=0.6\linewidth]{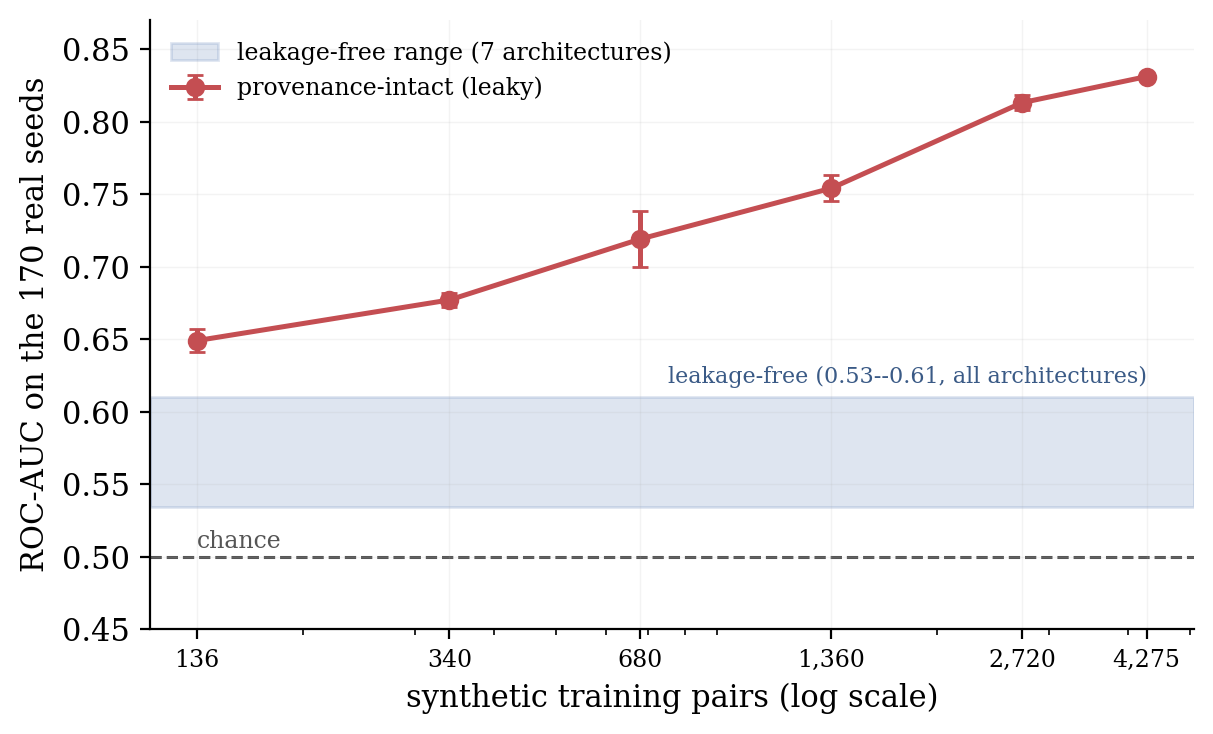}
  \caption{Random Forest ROC-AUC versus synthetic training-set size, with and without provenance overlap between training and test.}
  \label{fig:size-sweep}
\end{figure}

\runin{Why provenance control matters.}
Removing the provenance filter produces a large apparent performance increase. A Random Forest trained on synthetic pairs and evaluated on the real seeds from which those pairs were generated reaches ROC-AUC $0.831$ (Table~\ref{tab:real-vs-synth}). Performance remains $0.81$ even when the overlapping training set is reduced to $2{,}720$ pairs. This occurs despite the absence of exact seed reproduction and a maximum observed token 5-gram Jaccard similarity of only $0.448$ (Section~\ref{sec:dist:seed-distance}). Provenance overlap can therefore inflate performance without exact copying. We treat the $0.831$ result as a leakage diagnostic, not as a competing estimate of isolated-method performance. The provenance-uncontrolled result and its size-sweep points are retained as historical aggregates; their original prediction-level outputs are unavailable for independent recomputation. In contrast, the provenance-controlled results can be recomputed from the released out-of-fold predictions. Figure~\ref{fig:size-sweep} shows how this separation changes with synthetic training-set size.

\begin{finding}{3}
\emph{Under the models and evaluation settings tested here, no evaluated approach achieves strong and balanced isolated-method discrimination. Seven supervised architectures reach ROC-AUC $0.55$--$0.61$, and real-to-real training reaches approximately $0.64$. Zero-shot frontier models reach about $69\%$ balanced accuracy, but the two higher-performing probes correctly classify only $41.2\%$ of repaired methods as non-defective. This is an empirical result for the evaluated settings, not evidence that the method body contains no usable information.}
\end{finding}


\subsection{Pair Discrimination: Transferable Signal and the Repair-Growth Confound (RQ4)}
\label{sec:contrastive}

\subsubsection{Motivation and Task Formulation}

RQ4 changes the information available to the model while keeping the synthetic corpus, leave-seeds-out provenance control, and real evaluation population fixed. It uses its own seeded, source-bug-grouped folds, which are not exception-stratified (Section~\ref{sec:leave-seeds-out}). Unlike RQ3, the model can now compare the defective method directly with its developer repair.

In the \emph{pair-discrimination} task, the model receives both members of a bug/fix pair in randomised order and identifies the defective (pre-repair) method. This is a two-alternative forced choice (2AFC), so chance accuracy is $50\%$. The setting corresponds to cases where a candidate repair is already available and asks whether the pair provides transferable information about which version is defective.

Each pair is encoded as
\[
\texttt{[CLS]}\; m_A \;\texttt{[SEP][SEP]}\; m_B \;\texttt{[SEP]},
\]
and both orderings are used during training and evaluation. This makes the positional label distribution exactly balanced and ensures that reported accuracy is the mean of the two position-conditional accuracies rather than a position shortcut.

No real pair is used for training. The experiment therefore tests whether pair-discriminative signal learned only from RXSynth transfers to unseen real production defects.
 
\subsubsection{Experimental Protocol}
\label{sec:utility:protocol}

\runin{Models and training.}
We fine-tune CodeBERT~\cite{feng2020codebert} (\texttt{microsoft/codebert-base}, $125$M parameters) and CodeT5~\cite{wang2021codet5} (\texttt{Salesforce/codet5-base}, $220$M) under the same pair-classification protocol. Both use a sequence-classification head and a $512$-token input budget. Training uses AdamW for six epochs with learning rate $2\times10^{-5}$ and an effective batch size of $16$; the remaining optimisation settings are provided in the replication package. Each training pair is presented in both method orderings.

We use the final-epoch model rather than selecting a checkpoint on real validation data. The real pairs form the held-out evaluation set and also supply the seeds used for generation, so selecting checkpoints on those pairs would weaken the intended synthetic-to-real separation. All training settings are therefore fixed before evaluation.

\runin{Leakage control.}
For each run seed, the RQ4 experiment constructs source-bug folds in memory rather than loading precomputed fold files. The archived fold assignments were subsequently regenerated from the run seeds and verified against saved predictions. RQ4 applies the same source-bug-level provenance exclusion as RQ3, despite their different fold-construction procedures. Depending on the RQ4 fold and run seed, $2{,}573$--$3{,}180$ synthetic pairs remain for training.

\runin{Convergence gate.}
Occasionally a randomly initialised classification head remains near the random-guess plateau. We re-initialise and retrain a fold when its final-epoch \emph{training} loss exceeds $0.65$, allowing at most five attempts. The gate uses training loss only and never inspects test predictions, so it cannot select a model based on held-out performance. Under this rule, all $25$ folds used in the reported runs converged.
 
\subsubsection{Statistical Analysis}
\label{sec:utility:stats}

The primary null hypothesis is chance pair discrimination, $H_0:\pi=0.5$. Each real pair is evaluated in both method orderings, so the two predictions from a pair are dependent. We therefore treat the pair, rather than the individual ordering, as the unit of independence. For each run seed, we compute a $95\%$ interval using a pair-cluster bootstrap with $10{,}000$ resamples of the $170$ real pairs while keeping both orderings together. A run seed is above chance when the interval's lower bound exceeds $0.5$. We also report Cohen's $h$ relative to chance as an effect-size summary.

Changing the run seed re-partitions the source-bug folds as well as re-initialising training: between $132$ and $135$ of the $170$ real pairs move to a different fold across the five seeds. The across-seed dispersion therefore reflects the combined variation induced by the seeded partition and the training protocol, not optimisation variability alone, and it is distinct from sampling variability in the fixed real evaluation set. Across seeds we therefore report the mean accuracy, sample standard deviation, and a $t$-based $95\%$ interval on the mean, separately from the pair-cluster inference above. Per-type results are treated as exploratory: six of the seven exception types contain fewer than $25$ independent real pairs, so we report intervals but do not perform type-wise hypothesis tests.
 
\subsubsection{Results}
\label{sec:utility:results}

\begin{table}[tbp]
  \centering
  \caption{CodeBERT pair-discrimination accuracy on the $170$ real pairs, by run seed.}
  \label{tab:pair-seeds}
  \small
  \setlength{\tabcolsep}{6pt}
  \begin{tabular}{lcc}
    \toprule
    Seed & Accuracy & 95\% CI \\
    \midrule
    42   & 71.2\% & [67.1, 75.3] \\
    7    & 71.2\% & [67.4, 75.0] \\
    123  & 71.5\% & [67.6, 75.3] \\
    2024 & 69.4\% & [65.3, 73.5] \\
    555  & 68.5\% & [64.7, 72.4] \\
    \midrule
    Mean $\pm$ sd           & $70.4 \pm 1.3$ & [68.7, 72.0] \\
    Cohen's $h$ vs.\ chance & $0.42$         & \\
    Chance                  & $50.0\%$       & \\
    \bottomrule
  \end{tabular}
\end{table}

Models trained only on synthetic pairs transfer pair-discriminative signal to unseen real defects. CodeBERT reaches a mean accuracy of $70.4\%$ over five run seeds (sd $1.3$; $95\%$ interval on the mean $[68.7,72.0]$), compared with $50\%$ chance (Table~\ref{tab:pair-seeds}). The pair-cluster bootstrap interval excludes chance for every run seed. CodeT5 reaches $74.8\%$ overall ($95\%$ CI $[74.3,75.4]$; Table~\ref{tab:pair-baselines}). No real pair enters training, and synthetic descendants of the held-out seeds are removed by the provenance filter.

This establishes transfer under the pair formulation, but not the source of the signal. The repair introduces directional differences between the two methods that may support discrimination without identifying the underlying defect mechanism. We therefore test simple surface baselines before interpreting the encoder results.

\begin{table}[tbp] 
  \centering
  \caption{CodeBERT pair-discrimination accuracy by exception type, pooled over five seeds. Truncated: mean share of each real pair's tokens beyond the $512$-token input budget.}
  \label{tab:pair-types}
  \small
  \setlength{\tabcolsep}{5pt}
  \begin{tabular}{lrccc}
    \toprule
    Exception type & Pairs & Accuracy & 95\% CI & Truncated \\
    \midrule
    \texttt{NullPointerException}            &  87 & 77.0\% & [72.3, 81.8] & 34.8\% \\
    \texttt{IllegalArgumentException}        &  24 & 58.3\% & [53.3, 64.2] & 45.5\% \\
    \texttt{ArrayIndexOutOfBoundsException}  &  20 & 65.5\% & [57.0, 74.5] & 49.0\% \\
    \texttt{NumberFormatException}           &  13 & 50.0\% & [50.0, 50.0] & 75.8\% \\
    \texttt{ClassCastException}              &   9 & 82.2\% & [68.9, 94.4] & 37.6\% \\
    \texttt{StringIndexOutOfBoundsException} &   9 & 63.3\% & [46.7, 80.0] & 71.1\% \\
    \texttt{IllegalStateException}           &   8 & 73.8\% & [58.8, 88.8] & 27.8\% \\
    \bottomrule
  \end{tabular}
\end{table}

The per-type estimates are exploratory. Only \NPE has more than $25$ real pairs ($87$); each of the other six types has fewer than $25$, producing estimates that are too sparse for type-wise hypothesis testing. We therefore do not rank exception types or infer differences in their intrinsic difficulty from Table~\ref{tab:pair-types}. The next analysis instead tests whether aggregate performance is associated with directional repair cues.

\subsubsection{Surface Baselines and the Repair-Growth Confound}
\label{sec:utility:shortcut}

Simple surface baselines outperform both fine-tuned encoders. In the real evaluation set, the repaired method contains more tokens than the defective method in $78.8\%$ of pairs, the two have equal length in $11.8\%$, and the repaired method is shorter in only $9.4\%$. A parameter-free rule that labels the longer method as repaired and assigns half credit to ties therefore reaches $84.7\%$ expected accuracy ($95\%$ CI $[79.7,89.4]$). Logistic Regression and Random Forest models trained on synthetic differences in the $35$ structural and lexical features of Section~\ref{sec:fidelity} reach $82.1\%$ and $83.2\%$ on the real pairs (Table~\ref{tab:pair-baselines}). These values exceed CodeBERT's $70.4\%$ and CodeT5's $74.8\%$. CodeBERT is $11.7$ points below the synthetic-trained Logistic Regression model (pair-cluster bootstrap $95\%$ CI $[+6.2,+16.9]$). A feature model trained and evaluated on the real pairs reaches $85.6\%$. Aggregate pair accuracy therefore contains a strong surface component and should not by itself be interpreted as semantic defect understanding.

\runin{Residual no-growth stratum.}
The trained-model result weakens in the subset without positive repair growth. Among the $36$ real pairs where the repair is not longer than the defective method, the length rule falls to $27.8\%$ by construction and the synthetic-trained feature models fall to $45.8\%$ and $34.7\%$. CodeBERT reaches $54.2\%$ $[50.0,59.2]$ and CodeT5 $50.0\%$ $[41.9,58.3]$, with both intervals reaching $50\%$. Their paired difference is $-4.2$ points $[-10.3,+2.5]$ and spans zero, although CodeT5 is $4.5$ points higher overall ($74.8\%$ versus $70.4\%$; paired difference $[+1.5,+7.5]$). The overall CodeT5 advantage is therefore not evident on this subset, where neither encoder is distinguishable from chance or from the other. These results are consistent with much of the trained encoders' aggregate performance being associated with directional repair cues.

The near-chance encoder result does not mean that the residual pairs are indistinguishable. On the same $36$ pairs, GPT-5.5 identifies the defective member in $94.4\%$ of cases, Gemini~3.1~Pro in $97.2\%$, and Llama~3.3~70B in $77.8\%$. These zero-shot reference points indicate that the residual pairs remain highly discriminable when the dominant repair-growth cue is absent. Section~\ref{sec:diff:2afc} examines these models further, including position bias and the synthetic comparison arm.

\runin{Truncation and the same cue.}
The $512$-token pair budget truncates $71.8\%$ of the real pairs. CodeBERT reaches $86.7\%$ $[81.5,91.5]$ on the $48$ pairs that fit within the budget and $63.9\%$ $[60.3,67.7]$ on the $122$ truncated pairs. This comparison is not causal because pair length and truncation are confounded. On the in-budget pairs, CodeBERT's $86.7\%$ is also close to the length rule's $87.5\%$. The pattern is therefore consistent with truncation weakening the repair-growth cue, rather than showing that more context would necessarily recover semantic signal. CodeT5 remains $+2.5$ points above CodeBERT on the $48$ pairs both models see whole and $+5.7$ points above it on the $91$ pairs both truncate, so its overall advantage cannot be attributed only to tokenizer-dependent context coverage.

\begin{table}[tbp]
  \centering
  \caption{Pair accuracy (\%) of surface baselines, trained encoders, and zero-shot probes on the real pairs; brackets give $95\%$ CIs.}
  \label{tab:pair-baselines}
  \small
  \begin{tabular}{llcc}
    \toprule
    Model & Training & All pairs & Repaired not longer ($n{=}36$) \\
    \midrule
    Length sign rule      & none                        & $84.7$ $[79.7, 89.4]$ & $27.8$ \\
    $\Delta$-feature LR   & synthetic, leave-seeds-out  & $82.1$ $[76.8, 87.1]$ & $45.8$ \\
    $\Delta$-feature RF   & synthetic, leave-seeds-out  & $83.2$ $[77.9, 88.2]$ & $34.7$ \\
    $\Delta$-feature LR   & real, five-fold CV          & $85.6$ $[80.9, 90.0]$ & --- \\
    CodeBERT              & synthetic, leave-seeds-out  & $70.4$ $[68.7, 72.0]$ & $54.2$ $[50.0, 59.2]$ \\
    CodeT5                & synthetic, leave-seeds-out  & $74.8$ $[74.3, 75.4]$ & $50.0$ $[41.9, 58.3]$ \\
    \addlinespace
    Llama~3.3~70B         & zero-shot                   & $81.8$ & $77.8$ \\
    GPT-5.5               & zero-shot                   & $98.2$ & $94.4$ \\
    Gemini~3.1~Pro        & zero-shot                   & $97.6$ & $97.2$ \\
    \bottomrule
  \end{tabular}
\end{table}

\subsubsection{Frontier Reference Points}
\label{sec:diff:2afc}

We use three contemporary language models as zero-shot reference points on the same $170$ real pairs. GPT-5.5 identifies the defective member in $98.2\%$ of pairs, Gemini~3.1~Pro in $97.6\%$, and Llama~3.3~70B in $81.8\%$ (Table~\ref{tab:2afc}). These are reference points rather than trained-corpus results; the models receive no labelled examples from RXSynth.

\runin{Residual discrimination without repair growth.}
The no-growth subset provides the more informative comparison. On the $36$ pairs where the repair does not lengthen the method, both fine-tuned encoders are near chance, while GPT-5.5 and Gemini~3.1~Pro remain above $94\%$ and Llama~3.3~70B reaches $77.8\%$. This indicates that the residual pairs remain discriminable beyond the dominant repair-growth cue measured here, even though the two fine-tuned encoders do not recover that signal. The result does not identify what representation or reasoning process the zero-shot probes use.

\runin{Synthetic-arm control.}
The same zero-shot probes also discriminate synthetic pairs at similar rates: GPT-5.5 reaches $96.7\%$ and Gemini~3.1~Pro $98.6\%$. The production generator is Claude Opus~4.7, so neither frontier probe belongs to the generator family. Although the corpus validator Gemini~2.5~Flash shares a family with Gemini~3.1~Pro, GPT-5.5 shares a family with neither the generator nor the validator. Its $96.7\%$ synthetic-arm result shows that high synthetic-pair discrimination also occurs for a probe that shares a family with neither model used to generate or validate the corpus.

\runin{Position bias.}
Llama~3.3~70B shows a strong first-option preference: on the real arm its two position-conditional accuracies are $97.8\%$ and $62.8\%$. Averaging those rates gives $80.3\%$, still above chance but lower than its raw result suggests. Its synthetic position-balanced rate is $82.6\%$. GPT-5.5 and Gemini~3.1~Pro show little corresponding asymmetry. Their high aggregate accuracy therefore shows little evidence of being driven by answer position, while Llama's result should be interpreted with its position bias in mind.

\begin{table}[tbp] 
  \centering
  \caption{Zero-shot pair-discrimination (2AFC) accuracy on real and synthetic pairs, one trial per pair. A / B: accuracy on pairs where the defective method is shown first / second; $p$: exact two-sided binomial test against $50\%$.}
  \label{tab:2afc}
  \small
  \setlength{\tabcolsep}{4pt}
  \begin{tabular}{llrlcc}
    \toprule
    Model & Arm & Acc.\ ($k/n$) & 95\% CI & $p$ & A / B \\
    \midrule
    GPT-5.5        & real      & 98.2 (167/170) & [94.9, 99.4] & $10^{-45}$ & 98.9 / 97.4 \\
    GPT-5.5        & synthetic & 96.7 (203/210) & [93.3, 98.4] & $10^{-51}$ & 95.3 / 98.1 \\
    Gemini~3.1~Pro & real      & 97.6 (166/170) & [94.1, 99.1] & $10^{-44}$ & 97.8 / 97.4 \\
    Gemini~3.1~Pro & synthetic & 98.6 (207/210) & [95.9, 99.5] & $10^{-57}$ & 98.1 / 99.0 \\
    Llama~3.3~70B  & real      & 81.8 (139/170) & [75.3, 86.8] & $10^{-17}$ & 97.8 / 62.8 \\
    Llama~3.3~70B  & synthetic & 82.9 (174/210) & [77.2, 87.4] & $10^{-23}$ & 97.2 / 68.0 \\
    \bottomrule
  \end{tabular}
\end{table}

\begin{finding}{4}
\emph{Pair discrimination transfers from synthetic training data to unseen real defects, but high pair accuracy is not by itself evidence of semantic defect understanding. A parameter-free repair-growth rule reaches $84.7\%$ expected accuracy, and synthetic-trained surface-feature models reach $82.1$--$83.2\%$, both outperforming the fine-tuned encoders. On the $36$ pairs where the repair does not lengthen the method, CodeBERT and CodeT5 return to approximately chance, with intervals including $50\%$. This is consistent with much of their aggregate performance being associated with directional repair cues. GPT-5.5 and Gemini~3.1~Pro remain above $94\%$ on the same residual pairs, indicating that those pairs remain highly discriminable beyond the dominant repair-growth cue measured here. Semantic interpretations still require controls for repair-direction shortcuts and other possible sources of signal.}
\end{finding}

\section{Dataset Availability and Intended Use}
\label{sec:availability}
RXSynth is intended as a reusable research resource, so the release makes both its contents and its limits explicit. This section describes the dataset partitions and record format, the accompanying reproduction materials, and the intended and out-of-scope uses.

\subsection{Distribution and Format}
\label{sec:avail:format}

RXSynth is archived on Zenodo~\cite{eslamloo_2026_22695176}. The generated corpus data and accompanying metadata are released under CC~BY~4.0, and the accompanying code under the MIT licence. The methods contained in the $170$ real seed pairs retain the licences of their original source projects, with source-project provenance preserved in the release. The evaluated release is versioned; subsequent corrections will appear as new Zenodo versions with changelogs. The core, excluded, and zero-shot contrast collections use JSONL, with one object per line. The real seed pairs are stored as JSON arrays, while per-pair generation metadata and the source-bug identifier map are CSV files. These collections have different record schemas (Table~\ref{tab:release-partitions}).

\begin{table}[tbp]
  \centering
  \caption{Partitions in the RXSynth release.}
  \label{tab:release-partitions}
  \small
  \setlength{\tabcolsep}{5pt}
  \begin{tabular}{@{}lp{0.43\linewidth}p{0.31\linewidth}@{}}
    \toprule
    Partition & Contents & Purpose \\
    \midrule
    Core
      & $4{,}275$ validated seed-grounded pairs
      & Main corpus \\
    Excluded
      & $625$ candidates rejected during validation
      & Audit and rejection analysis \\
    Contrast
      & $7{,}000$ zero-shot pairs ($1{,}050$ judged; $5{,}950$ unjudged)
      & Grounding comparison \\
    Seeds
      & $170$ real bug/fix pairs, including the $13$ upper-tail pairs excluded from the $157$-pair fidelity reference set
      & Provenance and reference \\
    \bottomrule
  \end{tabular}
\end{table}

Each core record contains identifiers (\texttt{pair\_id}, \texttt{sample\_id}), the exception label (\texttt{error\_type}), buggy and fixed method bodies, partition membership, and generation-exemplar provenance through \texttt{example\_bug\_ids}. Core records also retain the accepting judge's metadata, including the judge model, confidence, located throw site, and trigger. Structured judgment logs are distributed separately and keyed by \texttt{sample\_id}; the original raw validation-response text was not fully retained. Rejected records have a separate schema with a rejection category and, for judged candidates, the judge's stated reason; they do not retain exemplar provenance. Contrast records use \texttt{id} as their key and do not embed semantic-validity labels; judgments for the $1{,}050$ audited pairs are provided separately. Real seeds use \texttt{bug\_id} and \texttt{exception\_canonical}, and store code in \texttt{buggy\_method} and \texttt{fixed\_method}. The released schema documents each collection's field types and optional fields.

Two release flags affect dataset selection. The $405$ \ISE pairs are explicitly flagged because Section~\ref{sec:val:final} identifies lower validation reliability for this exception type, allowing studies that require uniform label confidence to exclude them. The \texttt{fix\_introduces\_label} field marks $79$ pairs whose fixed method introduces an explicit throw of the pair's labelled exception type. These pairs remain valid under the patch-local criterion of Section~\ref{sec:val:semantic}, but should be excluded from experiments that instead define correctness through existential class-level throwability.

\subsection{Datasheet}
\label{sec:avail:datasheet}

A datasheet following Gebru et al.~\cite{gebru2018datasheets}, covering motivation, composition, collection, preprocessing, intended uses, distribution, and maintenance, is included in the replication package.

\subsection{Reproduction Package}
\label{sec:avail:repro}

The replication package includes the generation scripts and their prompt-construction logic, recorded prompt hashes, judge prompt materials, execution-harness template, validation and repair scripts, classifier configurations, and source-bug fold assignments for Section~\ref{sec:usecase}. Append-only JSONL judgment logs preserve the accept/reject history of the validation pipeline. The shipped fold assignments were reconstructed from archived run seeds and verified against the saved predictions; the original training runs constructed their folds in memory rather than loading these files.

Because several stages use hosted language models whose behaviour may change across snapshots, the package preserves available generation responses and structured validation records. These materials support downstream reanalysis without new API calls. Model identifiers and timestamps are recorded where the historical runs preserved them, rather than claiming uniform coverage.

The released datasets, validation records, provenance mappings, and archived predictions support auditing and reanalysis of the reported results. However, some historical construction stages cannot be reconstructed end to end from the released artifacts alone. These include the original derivation of the bug-level seed-extraction counts, the original zero-shot generation implementation and complete candidate pool, the historical execution metadata of the real-seed causal-validation stage, and the original lexical-marker extractor. The retained code does identify the baseline \IAE\ validation-density gate and shows that the follow-up probe disabled it. These limitations affect reconstruction of parts of the original data-construction history, rather than the availability of the released records, provenance information, validation logs, or archived experimental outputs. The accompanying limitations document identifies the preserved evidence and the remaining gaps.

\subsection{Intended and Out-of-Scope Use}
\label{sec:avail:use}

RXSynth is intended for contrastive bug/fix research when both method versions are available, patch-oriented validation of candidate repairs, and studies of seed-grounded synthetic-data generation. Its provenance records and contrast partition also support analyses of how grounding choices affect synthetic-code distributions. For context-aware research, RXSynth's method-only pairs provide the baseline against which added caller, contract, or execution context can be measured.

The corpus is not intended as an isolated-method defect-detection leaderboard, because reliable isolated-method discrimination is not established under the settings evaluated in Section~\ref{sec:difficulty}. It should not be used to claim representativeness for substantially larger methods, for which Section~\ref{sec:preferred-band} identifies systematic compression. The dataset is restricted to method-contained defects and therefore does not represent multi-method or interprocedural faults. Studies requiring uniform validation confidence should also apply the \ISE reliability flag and \texttt{fix\_introduces\_label} according to their required label semantics.

\section{Discussion}
\label{sec:discussion}

The empirical study leaves two main issues to explain: why access to the repair changes discrimination, and what that means for method-level detection and synthetic-data design. We discuss the role of surrounding program context, the consequences for context and provenance, the difference between generation and detection, and the limits of seed-grounded synthesis.

\subsection{Why Contrast Carries the Signal}
\label{sec:disc:contrast}

One explanation consistent with the RQ3 and RQ4 results is that defect status often depends on information outside the method body. A nullable parameter is problematic only if callers can supply null. A missing precondition matters only if checking it is the responsibility of the method rather than its callers. An array access may also be safe if the index has already been constrained upstream. In each case, the operation visible inside the method can look risky without being enough to determine whether it violates the intended contract. An isolated method can therefore omit information that is relevant to deciding whether the observed code is defective.

The isolated zero-shot results are consistent with this explanation. The models are much better at identifying potentially risky methods than at correctly labelling repaired methods as non-defective. This pattern suggests that recognizing a possible failure path is easier than deciding whether that path represents an actual defect under the intended usage conditions. A visible repair changes the evidence available to the model. Guards, checks, changed conditions, or altered control flow can reveal what the developer intended to prevent, even when that intent was not explicit in the original method.

This interpretation is also consistent with prior observations that buggy code can remain locally natural~\cite{ray2016naturalness}, that static bug detection can produce substantial false-positive burdens~\cite{du2026reducing}, and that judgments over isolated snippets can be harder than direct comparisons~\cite{alrashedy2023language}. These connections do not establish that caller contracts are the missing variable in every case. They instead suggest a broader information problem: local code may expose a risky operation without exposing enough of its intended use to determine whether the operation is wrong. Pair comparison can partially reduce that uncertainty because the repair provides direct evidence of developer intent.

\subsection{Implications for Method-Level Bug Detection}
\label{sec:disc:implications}

The first implication is that future work should study the information supplied to the model, not only the capacity of the model itself. Under the settings tested in Section~\ref{sec:difficulty}, changing architecture, training source, and inference regime does not produce strong and balanced isolated-method discrimination. RQ4 shows that changing the available information can alter the task much more substantially. A developer repair is an unusually strong source of context and is not available when searching for previously unknown defects, but weaker sources can be studied systematically. These include caller and callee methods, call sites, contracts or parameter constraints, data-flow and static-analysis facts, and execution traces. A useful next step is therefore to add such context incrementally and measure which sources improve discrimination and how much context is needed.

This also changes how method-level benchmarks should be interpreted. A poor isolated-method result does not by itself show that a larger model is required; it may indicate that important evidence is missing from the input. Conversely, a strong pairwise result does not by itself show that the model has learned the defect mechanism, because the repair can introduce directional surface cues. Evaluation should therefore vary the available information explicitly and include simple baselines that test whether performance can be explained by properties of the repair rather than by the underlying bug.

The second implication concerns provenance. Seed-grounded generation creates relationships between synthetic examples and the real examples used to produce them. Section~\ref{sec:leave-seeds-out} shows that these relationships matter even without exact copying. Similar dependencies can arise when training and test examples share source bugs, projects, patches, templates, or generation seeds. For generated or mined benchmarks, provenance should therefore be treated as part of the evaluation design rather than as metadata recorded only for reproducibility. Reporting the unit of provenance and how overlap is prevented makes it easier to distinguish transfer to an unseen defect from recognition of a related generation neighbourhood.

\subsection{The Generation--Detection Asymmetry}
\label{sec:disc:asymmetry}

Generation and isolated detection receive different information. During seed-grounded generation, the intended failure is part of the conditioning. The target exception is specified, and real bug/fix examples show both the defect pattern and the kind of repair that addresses it. Isolated detection starts from the opposite direction. The model sees a method and a target exception but must decide whether a potentially risky operation is actually unintended, often without the caller behaviour or surrounding contract that would resolve that question. The results therefore reveal an information asymmetry between the two tasks. They do not establish that generation is intrinsically easier than detection in general.

The same distinction helps explain why patch-anchored validation can be useful even when unaided isolated detection remains difficult. The validator does not need to discover an arbitrary defect from scratch. It first identifies the risk associated with the changed location and then asks whether the repair removes that specific risk. The blinded human audit in Section~\ref{sec:val:robustness} exposed ambiguity in whole-method judgments and motivated this patch-local criterion. The criterion is then evaluated through comparison across three LLM judges from two model families and corroborated through independent JVM execution. This sequence is important: the human audit motivates the question being asked, while the cross-judge comparison and execution provide evidence about the resulting criterion.

The broader point is that model capability cannot be separated from task formulation. Asking whether a method is globally safe is different from asking whether a particular patch removes a particular risk. A model may therefore be useful for structured validation when the relevant site, change, and target exception are available even if the same model family is unreliable at discovering defects from an isolated method.

\subsection{Limitations of Distribution-Matched Synthetic Generation}
\label{sec:disc:limitations-of-synthetic}

Seed grounding improves similarity to the examples it receives, but it cannot guarantee realistic code outside that range. Section~\ref{sec:preferred-band} shows that generated methods follow their conditioning exemplars more closely within the structural range that the generator reproduces well, while larger exemplars are systematically compressed. The important consequence is not simply that some generated methods are shorter. Structural regions that are absent from the seeds, or that the generator does not reproduce faithfully, remain poorly represented even if many more synthetic descendants are generated.

Grounding also transfers some properties more easily than others. The fidelity analysis in Section~\ref{sec:dist:token} shows closer agreement for constructs directly tied to the exception mechanism than for several incidental constructs that surround those defects in production methods. This suggests that an exemplar can communicate the failure pattern more reliably than the broader program context in which that pattern originally appeared. A generated method may therefore reproduce a realistic null check, bounds error, or invalid-state pattern while still simplifying the unrelated work performed around it. This distinction matters because downstream models can learn from both the intended defect mechanism and the synthetic context that surrounds it.

These results imply that synthetic scale and source diversity are different resources. Generating more descendants from the same seeds can increase sample count and combine existing patterns in new ways, but it cannot by itself create structural regimes that are missing from the conditioning data or not reproduced by the generator. For extending coverage, the results therefore suggest collecting more real seeds from under-represented method sizes, projects, APIs, and programming contexts before simply increasing the number of generations from the existing pool.

This is a broader lesson for synthetic-data construction. Grounding is most useful as a controlled way to interpolate around observed examples, not as evidence that the resulting corpus represents unrestricted production code. Increasing synthetic volume, improving semantic validity, and broadening distributional coverage are separate goals. A larger corpus can satisfy the first two while still missing important parts of the target population. Future extensions of RXSynth should therefore treat seed coverage and generator fidelity as first-class design choices alongside corpus size.
\section{Threats to Validity}
\label{sec:threats}
This study has several limits that affect how the results should be interpreted. We group them into construct, internal, external, and conclusion validity. The main concerns are the patch-local validation target, dependence on model-based judging and repair, provenance and truncation, the scope of the Java seed pool, possible pretraining contamination, and uncertainty from the limited real sample.

\subsection{Construct Validity}
The pair task measures discrimination between the defective and repaired members of a developer patch, not existential throwability and not defect detection from an isolated method. A repaired method may legitimately throw the same exception type on another input or at another site while still being the repaired member of the pair. Our validation criterion is therefore patch-local: it asks whether the buggy method exposes the labelled risk at the patched site and whether the repair removes that risk on the same trigger. This definition matches the corpus's intended pair-discrimination use but should not be substituted for whole-method safety. The release flags in Section~\ref{sec:avail:format} make this boundary explicit for analyses that require different semantics.

Validation is also not equivalent to a complete behavioural proof. The main pipeline relies on LLM judgments. A blinded human audit exposed ambiguity in whole-method judgments and motivated the patch-anchored criterion. We then evaluate that criterion through cross-judge comparison in Section~\ref{sec:val:robustness} and independent JVM execution in Section~\ref{sec:val:execution}. The execution harnesses themselves are model-generated and use stubs for unavailable dependencies; although every harness is compiled and run, we cannot guarantee that each stub reproduces the behaviour of the original production environment. Some valid pairs may therefore be unverifiable under the harness constraints, while some exercised paths may omit behaviours present in the real project. The evidence supports patch-local validity under the stated criterion, not semantic equivalence of the repaired method to its production counterpart or correctness for all possible inputs.

\subsection{Internal Validity}
Training reproducibility was measured rather than assumed. Repeating runs with identical data, identical seeds and an identical environment, CodeT5, TextCNN, BiLSTM and XGBoost produced bit-identical predictions. The Random Forest's per-row probabilities varied, but its reported ROC-AUC reproduced to six decimal places, because the varying rows did not change the ranking. CodeBERT and CodeGPT showed genuine training nondeterminism: on folds whose training data was held byte-identical, pooled CodeBERT accuracy moved by up to $5.6$ percentage points, which exceeds the aggregate difference between our earlier and current CodeBERT pair results. Differences of that magnitude in CodeBERT should therefore not be read as effects of any change to the training set; in particular, we report CodeBERT's $70.4\%$ as the output of the current provenance-controlled pipeline and do not attribute its movement to the provenance filter. CodeT5 was deterministic under the same test, so its corresponding movement is attributable under the measured setup. These statements describe the runs reported here and are not general claims about these architectures.

The pair-training protocol includes a convergence gate that re-initialises a fold when final training loss remains above the prespecified threshold. The gate depends only on training loss and never on held-out predictions, so it does not select checkpoints using test performance; nevertheless, it makes convergence conditioning part of the reported training procedure. The repair loop has a different dependence: regenerated fixes are readmitted by the same production judge whose diagnosis prompted repair. Their readmission is therefore not an independent validation of each repaired item, and the cross-judge and execution analyses support the overall criterion rather than individually certifying every repaired pair.

Training/evaluation dependence is controlled through provenance and near-duplicate filtering. Generation-time similarity and clone checks reject near-duplicate synthetic outputs (Section~\ref{sec:gen:qc}), while synthetic descendants of held-out real seeds are removed fold by fold during evaluation. Section~\ref{sec:dist:seed-distance} additionally reports zero exact seed reproductions, with a maximum observed token 5-gram Jaccard similarity of $0.448$. These controls reduce direct copying and generation-neighbourhood recognition but cannot establish independence from all latent similarities. Pair encoding is also subject to a $512$-token limit; Section~\ref{sec:utility:shortcut} quantifies its association with performance and shows that the CodeT5 advantage persists within shared truncation strata. Finally, the pair-discrimination experiment evaluates two trained encoder architectures, so claims about trained pair models should not be generalized to all model families. 

\subsection{External Validity}
The corpus and empirical findings are scoped to Java and to the seven unchecked runtime-exception types retained by the seed-extraction process. Languages with different exception models, type systems, calling conventions, or coding idioms may exhibit different generation and detection behaviour. Likewise, the study does not cover checked exceptions or unchecked types too rare in the source benchmarks to support the same generation protocol.

The corpus is also deliberately method-contained. Filter F3 in Section~\ref{sec:filter-rules} excludes fixes whose relevant changes span multiple methods or files, so the results do not establish performance for interprocedural faults, state changes distributed across components, or repairs whose correctness depends on edits outside one method body. The observed compression of larger exemplars further limits representativeness for long methods, as discussed in Section~\ref{sec:preferred-band}. The $13$ upper-tail pairs are excluded from the fixed $157$-pair fidelity reference set, but production conditioning did not follow that boundary exactly: two \texttt{JacksonCore-19} pairs among them conditioned $82$ unique released descendants, while the other $11$ reference-excluded pairs contribute no released descendants. The generated core therefore covers the upper tail unevenly and should not be interpreted as representative of substantially larger methods. In addition, the real seeds come from four production-code bug benchmarks. They therefore reflect the projects, coding practices, and defect-reporting processes represented in those benchmarks rather than all Java software, and they should not be treated as representative of contest, educational, proprietary, or safety-critical code.

Pretraining contamination remains possible. The generator, fine-tuned encoders, and zero-shot probe models may have encountered the benchmark projects or related code during pretraining. Instance-level seed reproduction is not observed in Section~\ref{sec:dist:seed-distance}, and the leave-seeds-out protocol removes direct generation provenance from training to evaluation, but neither control can determine what a pretrained model memorized before our experiments. This caveat is especially relevant to the frontier zero-shot reference points, whose high pair accuracy should therefore be interpreted as an observed capability under the tested setting rather than as contamination-free generalization.

\subsection{Conclusion Validity}
Each real pair is evaluated in both method orderings, producing dependent predictions. We therefore treat the pair, not the individual ordering, as the unit of independence in interval estimation and hypothesis testing. The five run seeds in the pair experiment vary both the source-bug fold assignment and the training procedure, so across-seed dispersion combines partition and training variability rather than isolating optimisation variability.

The archived RQ4 folds are source-bug-grouped but not stratified by exception type. Although source-bug separation and provenance exclusion remain intact, the effect of using exception-stratified folds on the reported RQ4 estimates has not been measured. The results therefore describe the archived evaluation protocol rather than demonstrating robustness to alternative fold assignments.

These five runs do not provide enough independent evidence to support fine-grained claims about small differences between closely matched model configurations, and we avoid such comparisons. Per-exception analyses are likewise limited by the small number of real pairs for most types, so those results are reported descriptively and without type-wise hypothesis testing. More broadly, confidence intervals quantify uncertainty for the sampled benchmark pairs and training repetitions under the stated protocol; they do not remove the external-validity limits described above.

\section{Conclusion}
\label{sec:conclusion}

Existing Java bug benchmarks provide too few method-level runtime-exception bug/fix pairs for controlled training and transfer studies. Mining four established benchmarks leaves only $170$ usable real pairs across seven unchecked exception types. RXSynth extends this evidence base with $4{,}275$ seed-grounded buggy/fixed pairs and records source-bug provenance for its generation exemplars. Its construction combines generator screening, patch-anchored semantic validation, and provenance tracking. A blinded human audit motivated the patch-anchored criterion, which we then evaluated through cross-judge comparison and independent JVM execution. Fidelity analysis also defines the corpus's limits: RXSynth is structurally close to the fixed $157$-pair fidelity reference set within the represented size range, while pair-level exemplar analysis shows increasing compression for larger methods and less faithful transfer of incidental program context. RXSynth should therefore be viewed as a controlled extension of the available real examples rather than a substitute for the full diversity of production Java.

The corpus also allows us to separate two different information settings. Under the models, training sources, and inference settings tested here, no evaluated approach achieves strong and balanced isolated-method discrimination. Making the repair visible changes the task: synthetic-trained pair classifiers transfer to unseen real defects, but much of the trained-model performance is associated with simple directional repair cues. A parameter-free repair-growth rule reaches $84.7\%$ expected accuracy, higher than either fine-tuned encoder. On the $36$ pairs where the repair does not increase token count, both encoders return to approximately chance. Given both methods, GPT-5.5 and Gemini~3.1~Pro remain above $94\%$ on the same residual pairs, indicating that those pairs remain discriminable when the dominant repair-growth cue is absent. These results show why high forced-choice accuracy on developer bug/fix pairs should not be interpreted as semantic defect understanding without simple directional baselines and residual analyses. They also motivate using RXSynth mainly for contrastive, patch-oriented, and context-aware studies rather than as an isolated-method detection leaderboard.

Three directions follow from these findings. First, \emph{context-aware detection} should test what information can substitute for an unavailable repair. Caller and callee code, call sites, contracts, data-flow facts, static-analysis results, and execution evidence can be added incrementally to measure which sources improve discrimination. Second, \emph{models beyond repair shortcuts} should be evaluated on settings where simple directional cues are weak or absent, rather than only on aggregate pair accuracy. Progress should include performance beyond length and surface-feature baselines on provenance-disjoint real defects. Third, \emph{richer synthesis} should expand the diversity of real seeds and improve fidelity for larger methods and surrounding program context instead of only generating more descendants from the current pool. More broadly, synthetic scale, semantic validity, and distributional coverage are separate properties. Benchmark design should treat provenance, fidelity, and available information as first-class concerns so that improved performance can be attributed more confidently to the capability being studied.

\begin{acks}
This research was supported in part by the Natural Sciences and Engineering Research Council of Canada (NSERC) Discovery Grants program, the Canada Foundation for Innovation's John R. Evans Leaders Fund (CFI-JELF), and the industry-stream NSERC CREATE in Software Analytics Research (SOAR) program.

The authors thank Richard Bassey and Raphael De Los Reyes for their contributions to the manual validation of the RXSynth dataset.
\end{acks}

\bibliographystyle{ACM-Reference-Format}
\bibliography{bibliography}

\end{document}